\documentclass[preprint]{aastex63}

\usepackage[fleqn]{amsmath}
\usepackage[hyphenbreaks]{breakurl}
\usepackage{float}
\usepackage{url}

\makeatletter
\newcommand*{\rom}[1]{\expandafter\@slowromancap\romannumeral #1@}
\makeatother

\accepted{Sep 5, 2023}
\submitjournal{ApJL}

\shorttitle{Multiband Period-Luminosity-Metallicity relations for Cepheid variables}
\shortauthors{Bhardwaj A. et al.}

\begin{document}
\title{High-resolution Spectroscopic Metallicities of Milky Way Cepheid Standards and their impact on the Leavitt Law and the Hubble constant}

\correspondingauthor{Anupam Bhardwaj}
\email{anupam.bhardwajj@gmail.com; anupam.bhardwaj@inaf.it}
\author[0000-0001-6147-3360]{Anupam Bhardwaj}\thanks{Marie-Curie Fellow}
\affil{INAF-Osservatorio Astronomico di Capodimonte, Via Moiariello 16, 80131 Napoli, Italy}
\author[0000-0002-6124-1196]{Adam G. Riess}
\affil{Space Telescope Science Institute, 3700 San Martin Drive, Baltimore, MD 21218, USA}
\affil{Department of Physics and Astronomy, Johns Hopkins University, Baltimore, MD 21218, USA}
\author{Giovanni Catanzaro}
\affil{INAF-Osservatorio Astrofisico di Catania, Via S.Sofia 78, 95123, Catania, Italy}
\author{Erasmo Trentin}
\affil{Leibniz-Institut für Astrophysik Potsdam (AIP), An der Sternwarte 16, D-14482 Potsdam, Germany}
\affil{Institut für Physik und Astronomie, Universität Potsdam, Haus 28, Karl-Liebknecht-Str. 24/25, D-14476 Golm (Potsdam), Germany}
\affil{INAF-Osservatorio Astronomico di Capodimonte, Via Moiariello 16, 80131 Napoli, Italy}
\author{Vincenzo Ripepi}
\affil{INAF-Osservatorio Astronomico di Capodimonte, Via Moiariello 16, 80131 Napoli, Italy}
\author[0000-0002-6577-2787]{Marina Rejkuba}
\affiliation{European Southern Observatory, Karl-Schwarzschild-Stra\ss e 2, 85748, Garching, Germany}
\author[0000-0002-1330-2927]{Marcella Marconi}
\affil{INAF-Osservatorio Astronomico di Capodimonte, Via Moiariello 16, 80131 Napoli, Italy}
\author[0000-0001-8771-7554]{Chow-Choong Ngeow}
\affil{Graduate Institute of Astronomy, National Central University, 300 Jhongda Road, 32001 Jhongli, Taiwan}
\author[0000-0002-1775-4859]{Lucas M. Macri}
\affil{NSF's NOIRLab, 950 N Cherry Ave, Tucson, AZ 85719}
\author{Martino Romaniello}
\affiliation{European Southern Observatory, Karl-Schwarzschild-Stra\ss e 2, 85748, Garching, Germany}
\author{Roberto Molinaro}
\affil{INAF-Osservatorio Astronomico di Capodimonte, Via Moiariello 16, 80131 Napoli, Italy}
\author[0000-0001-6802-6539]{Harinder P. Singh}
\affiliation{Department of Physics and Astrophysics, University of Delhi,  Delhi-110007, India }
\author{Shashi M. Kanbur}
\affiliation{Department of Physics, State University of New York, Oswego, NY 13126, USA}

\begin{abstract} 
Milky Way Cepheid variables with accurate {\it Hubble Space Telescope} photometry have been established as standards for primary calibration of the cosmic distance ladder to achieve a percent-level determination of the Hubble constant ($H_0$). These 75 Cepheid standards are the fundamental sample for investigation of possible residual systematics in the local $H_0$ determination due to metallicity effects on their period-luminosity relations. We obtained new high-resolution ($R\sim81,000$), high signal-to-noise ($S/N\sim50-150$) multi-epoch spectra of 42 out of 75 Cepheid standards using ESPaDOnS instrument at the 3.6-m Canada-France-Hawaii Telescope. Our spectroscopic metallicity measurements are in good agreement with the literature values with systematic differences up to $0.1$ dex due to different metallicity scales. We homogenized and updated the spectroscopic metallicities of all 75 Milky Way Cepheid standards and derived their multiwavelength ($GVIJHK_s$) period-luminosity-metallicity and period-Wesenheit-metallicity relations using the latest {\it Gaia} parallaxes. The metallicity coefficients of these empirically calibrated relations exhibit large uncertainties due to low statistics and a narrow metallicity range ($\Delta\textrm{[Fe/H]}=0.6$~dex). These metallicity coefficients are up to three times better  constrained if we include Cepheids in the Large Magellanic Cloud and range between $-0.21\pm0.07$ and $-0.43\pm0.06$ mag/dex. The updated spectroscopic metallicities of these Milky Way Cepheid standards were used in the Cepheid-Supernovae distance ladder formalism to determine $H_0=72.9~\pm 1.0$\textrm{~km~s$^{-1}$~Mpc$^{-1}$}, suggesting little variation ($\sim 0.1$ ~km~s$^{-1}$~Mpc$^{-1}$) in the local $H_0$ measurements due to different Cepheid metallicity scales. \\
\end{abstract} 

\section{Introduction}

Classical Cepheids play a vital role in calibrating the first-rung of the cosmic distance ladder thanks to their well-known Period-Luminosity (PL) relation  \citep[or the ``Leavitt Law'',][]{leavitt1912}, which allows an accurate and precise measurement of the Hubble constant ({\it $H_0$}) in the local universe \citep{freedman2001, riess2016, riess2022}. At present, a $\sim$1.4\% measurement of {\it $H_0$} in the late universe based on Cepheids and Type Ia Supernovae (SNe) is significantly ($\sim$5.3$\sigma$) larger than the precise {\it Planck} inference based on observations of the early universe \citep{planck2020, riess2022a}. This discrepancy between direct and cosmology-based $H_0$ measurements may lead to new physics beyond the standard model \citep{valentino2021}. Reaching an accurate, percent-level measurement of {\it $H_0$} while investigating all possible sources of systematic uncertainties in Cepheid-based measurements is now critical considering this Hubble tension \citep[see][for details]{verde2019}.

In the traditional Cepheid-SNe distance ladder from the SH0ES project \citep[Supernovae and $H_0$ for the equation of state of dark energy;][]{riess2011, riess2016, riess2022}, the main Cepheid-based error budget on $H_0$ comes from three anchors with geometric distances but receives the highest weight from {\it Gaia} Milky Way (MW) parallaxes \citep{riess2022}. Since the calibration of the Galactic Cepheid PL relation is now approaching percent-level precision \citep{riess2022a, reyes2023}, it is crucial to investigate systematic uncertainties in these empirical calibrations due to metallicity \citep[][]{riess2021, ripepi2021, breuval2022, molinaro2023,owens2022}. Specifically, given that the Cepheid-only error budget in $H_0$ measurements is below $1\%$ \citep{riess2022}, even modest differences due to past, heterogeneous metallicities or metallicity scales can challenge the goal of achieving the ultimate percent-level precision. 

High-resolution spectroscopic metallicities for several bright solar neighbourhood Cepheids are available in the literature that were primarily used to investigate the abundance gradient in the thin disk and/or the impact of metallicity on PL relations \citep[e.g.,][]{genovali2014, genovali2015, luck2018, ripepi2021, dasilva2022, kovtyukh2022}. Complementing metal-rich Cepheids in the solar vicinity, \citet{trentin2023} provided high-resolution spectra for the most metal-poor Galactic Cepheids that will be useful for an improved quantification of the metallicity dependence of PL relations. Some of these aforementioned studies also used multi-epoch spectra for a smaller sample of Cepheids to homogenize the metallicity scale of a larger sample of literature measurements \citep[e.g.,][]{genovali2014, genovali2015, luck2018}. The literature compilations of a few hundred Cepheids allow better statistical constraints on the metallicity dependence of PL relations, but cannot provide a proper account of systematic uncertainties due to heterogeneous metallicities obtained by employing diverse methodologies on spectra from different instruments/telescopes in different spectral ranges. Although there are more than 600 MW Cepheids with high-resolution spectra \citep{trentin2023}, only 75 bright nearby Cepheids have both the high-resolution spectroscopy and accurate {\it Hubble Space Telescope (HST)} photometry \citep{riess2021}.

\citet{riess2018, riess2021} established 75 MW Cepheid standards with {\it HST} photometry for calibration of the first-rung of the distance ladder. The {\it HST} photometry for these bright Cepheids was carried out in an observationally rigorous and specifically designed spatial scanning mode to mitigate saturation and reduce pixel-to-pixel calibration errors \citep{riess2018}. The {\it HST} photometry played a crucial role to establish the set of MW Cepheid standards calibrated on the same photometric system used to measure their extragalactic counterparts. The photometric calibration minimized systematic uncertainties due to different instrumental calibrations and/or ground-to-space based photometric transformations. However, the spectroscopic data used by \citet{riess2021} for the calibration of the $W^H_m$ Period-Wesenheit-Metallicity (PWZ) relation in {\it HST} filters come from a compilation of literature measurements. We have therefore started an observational project to collect high-resolution and high signal-to-noise spectra obtaining homogeneous spectroscopic metallicities for all these MW Cepheid standards. Our goal is to derive multiband Period-Luminosity-Metallicity (PLZ) relations and investigate possible small systematic uncertainties in the local $H_0$ determination due to metallicity dependence of Cepheid PL relations.   

\section{The Spectroscopic and photometric data}
\label{sec:data}

\begin{figure*}
\centering
\includegraphics[width=0.96\textwidth]{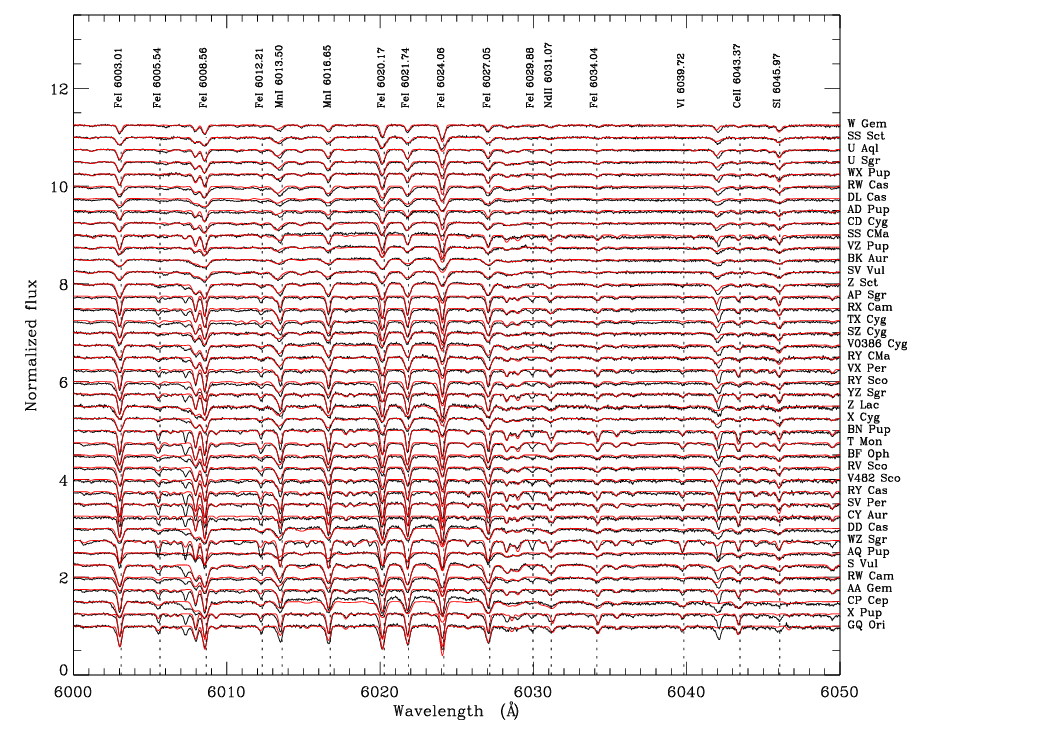}\\
\caption{High-resolution ESPaDOnS spectra for all 42 MW Cepheid standards in the range 6000-6050 \AA.  The star IDs are provided on the right and the vertical dashed lines mark the positions of atomic lines.} 
\label{fig:cfht_spec}
\end{figure*}

\subsection{New high-resolution spectroscopic metallicities}

 We obtained homogeneous high-resolution ($R\sim81,000$), high signal-to-noise ($S/N\sim 50-150$) optical spectra ($370~\textrm{to}~1050~ \textrm{nm}$) for 42 Cepheids that are observable with the ESPaDOnS instrument at 3.6-m Canada-France-Hawaii Telescope (CFHT). The observations were carried out in {\it star-only} mode and up to three spectra were obtained for a given target on 13 different nights across six runs between February 2021 and September 2022. A total of 90 pre-processed spectra (bias subtraction, spectrum extraction, flat fielding and wavelength calibration) were generated with the Libre-ESpRIT{\footnote{\url{https://www.cfht.hawaii.edu/Instruments/Spectroscopy/Espadons/Espadons_esprit.html}}} data reduction package for reducing echelle spectropolarimetric data at CFHT. 

The iron abundances were measured using the methodology discussed in detail in \citet{ripepi2021} and \citet{trentin2023}. In brief, the first step was to measure the atmospheric parameters --- effective temperature, surface gravity, microturbulent velocity and the line broadening parameter. The effective temperature was measured using the line-depth ratio (LDR) method \citep{kovtyukh2000}. The microturbulent velocities, iron abundances, and gravities were then determined through an iterative procedure in which it was assumed that the iron abundances do not depend on the equivalent widths. The equivalent widths of a sample of 145 Fe~{\footnotesize \rom{1}} lines published by \citet{romaniello2008} were used for this purpose. The equivalent widths were converted to abundances through the WIDTH9 code \citep{kurucz1993} applied to the corresponding atmospheric model calculated using ATLAS9 \citep{kurucz2005}. The influence of $\log g$ was neglected since neutral iron lines are insensitive to this parameter. For the surface gravities, an ionization equilibrium was imposed between Fe~{\footnotesize \rom{1}} and Fe~{\footnotesize \rom{2}}, where the line list for the latter was provided by \citet{romaniello2008}. 
The atmospheric parameters were used to compute stellar atmosphere models and synthetic spectra to analyze our observed spectra \citep[see][for details]{ripepi2021}. 
For the iron abundance, we obtained a weighted average of the values derived from the individual spectra available for each target. Fig.~\ref{fig:cfht_spec} displays random-phase spectra for each of the 42 MW Cepheids observed with CFHT. The details regarding the spectral observations, atmospheric and stellar parameters, and abundance determinations will be presented in Catanzaro et al. (2023, in prep.) together with new high-resolution near-infrared spectra for some of these Cepheid standards obtained using the IGRINS instrument at the 8-m Gemini South Telescope.

\begin{figure*}
\centering
\includegraphics[width=0.9\textwidth]{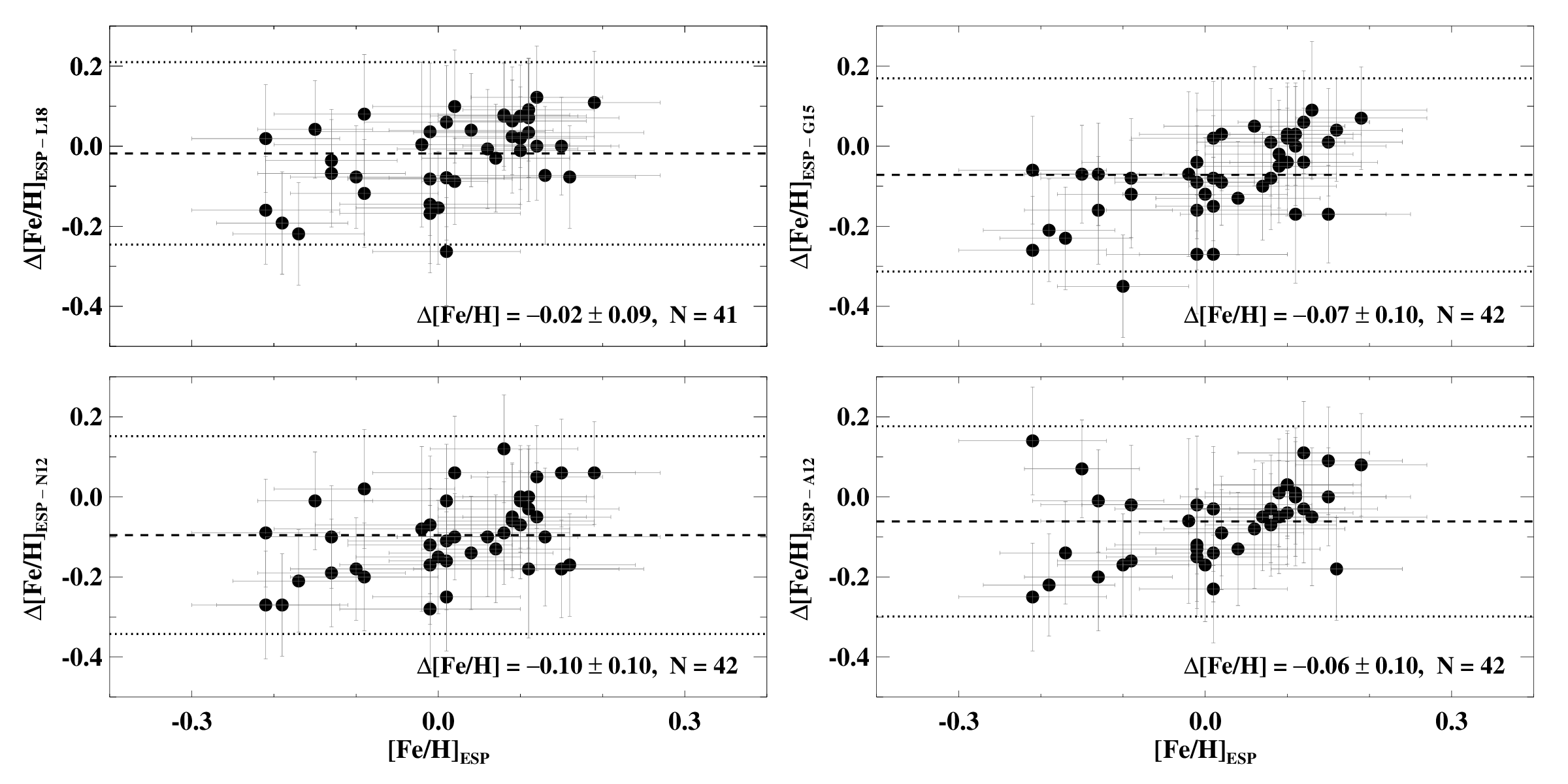}\\
\caption{Comparison of our [Fe/H]$_\textrm{ESP}$ values with measurements from \citet[L18,][]{luck2018}, \citet[G15,][]{genovali2015}, \citet[N12,][]{ngeow2012}, and \citet[A12,][]{acharova2012}. The mean offsets, standard deviation, and the number of common stars are listed on the bottom right of each panel. The dashed and dotted lines represent the mean and $\pm2.5\sigma$ offsets, respectively.} 
\label{fig:comp_feh}
\end{figure*}

\subsection{Comparison with literature metallicities}

We compared our spectroscopic iron abundances ([Fe/H]$_\textrm{ESP}$) with those available in the literature from different compilations. \citet{riess2021} used spectroscopic metallicities for all 75 MW Cepheid standards from the homogenized compilation by \citet{groenewegen2018}, which were originally either derived or homogenized by \citet{genovali2014, genovali2015} in their metallicity scale. We note that \citet{genovali2014} obtained high-resolution spectra for 42 Cepheids and combined their measurements with other literature values to derive a common metallicity scale for a sample of 450 Cepheids. We find that 74 Cepheids have [Fe/H] values in \citet{groenewegen2018} or \citet{genovali2014, genovali2015}, but CR Car is not included in these samples. \citet{riess2021} adopted a value of $-0.08$~dex for its iron abundance. Recently, \citet{kovtyukh2022} derived a [Fe/H] value of $0.14\pm0.07$~dex for CR Car, while \citet{trentin2023} found their measurements systematically overestimated by $+0.06$ dex. Therefore, we adopted a corrected [Fe/H] value of $0.08\pm0.08$ for CR Car. The updated sample\footnote{The [Fe/H] values of 11 Cepheids (V0340 Ara, S Tra, XX Car, V0482 Sco, V0339 Cen, XZ Car, DR Vel, SS CMa, AG Cru, XY Car, AQ Car) out of 75 in \citet{riess2021} were misattributed as small numbers ($<0.03$) due to an inversion between the value and uncertainty columns in \citet{groenewegen2018}. These were replaced with their true [Fe/H] values from \citet{genovali2014, genovali2015}. The impact on $H_0$ was an increase by 0.02 km/s/Mpc.} of \citet{riess2021} will be discussed as \citet{genovali2015} sample for the subsequent analysis.

Fig.~\ref{fig:comp_feh} displays a comparison of our [Fe/H]$_\textrm{ESP}$ values for 42 stars with the compilations by \citet{luck2018}, \citet{genovali2015}, \citet{ngeow2012}, and \citet{acharova2012}. The [Fe/H] values from these studies were adopted from the compilation of \citet{groenewegen2018} except for \citet{luck2018} measurements. The latter does not provide [Fe/H] values for RY CMa, and therefore, only 41 stars were considered for comparison. Our [Fe/H] measurements are in excellent agreement ($\Delta$[Fe/H]$=-0.02$~dex) with those from \citet{luck2018} based on multi-phase spectra of Cepheid variables. The largest difference is noted with the compilation by \citet{ngeow2012}, which is based on the metallicity scale of \citet{luck2011}. This was expected since \citet{luck2011} found a systematic offset of $+0.07$~dex in their metallicity measurements with previous literature estimates by their group, and a larger offset of $+0.11$~dex with independent measurements from \citet{romaniello2008}. These [Fe/H] offsets were attributed to higher surface gravity estimates ($\Delta \log g\sim+0.1$ leads to $\Delta$[Fe/H]$\sim+0.05$~dex) in \citet{luck2011}, which likely resulted from a subjective line editing of a smaller number of Fe~{\footnotesize \rom{2}} lines during the aforementioned iterative process. 

\citet{luck2018} reanalysed multi-phase spectra of Cepheids including those in the sample of \citet{luck2011} and found a small median offset of $\Delta\textrm{[Fe/H]}=-0.03$~dex with 
the metallicity scale of \citet{genovali2015}. Our $\textrm{[Fe/H]}_\textrm{ESP}$ values, while being consistent with \citet{luck2018}, are also systematically smaller by $\Delta$[Fe/H]$= -0.07$~dex than the measurements from \citet{genovali2015}. The largest difference of $\Delta$[Fe/H]$=0.35$~dex is noted for GQ Ori. However, \citet{genovali2015} mentions that the S/N of spectra for GQ Ori was not good enough for abundance measurements, which may be contributing to its relatively larger metallicity estimate as compared to our values and those in \citet{ngeow2012} or \citet{acharova2012}. The differences between [Fe/H] values from different studies are systematically larger for metal-poor stars which could arise due to small differences in effective temperatures. This is possible since the number of spectral line pairs is often not enough for LDR method to determine very accurate temperatures for these lower metallicity stars \citep[][]{trentin2023}. Nevertheless, the median uncertainty on our measurements is $0.09$~dex, and therefore, the offsets with literature compilations shown in Fig.~\ref{fig:comp_feh} are within $1\sigma$ of systematic uncertainties ($\sim 0.1$~dex) on spectroscopic metallicities in different scales/compilations. We also note that the [Fe/H] values of our 42 Cepheids from these literature compilations are not completely independent. Often the same literature measurements were homogenized by adopting a fixed zero-point in different studies, and therefore a detailed investigation into systematic differences is beyond the scope of this paper.  

\begin{figure}
\centering
\includegraphics[width=0.45\textwidth]{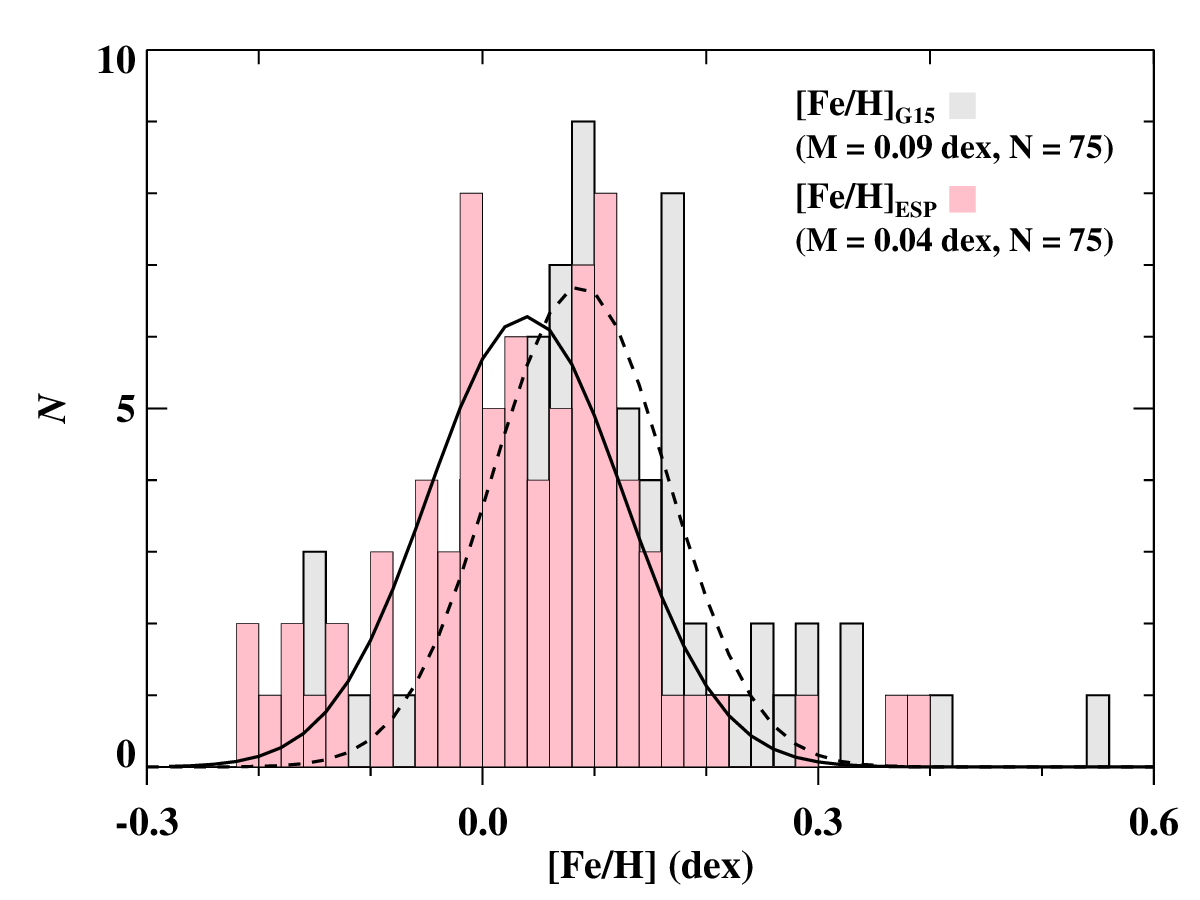}\\
\caption{Histograms of metallicities (in grey) for Cepheids from \citet{genovali2015} that were used in  \citet{riess2021}. Our homogenized sample of [Fe/H]$_\textrm{ESP}$ values (in pink) for all 75 Cepheids are also overplotted. The solid and dashed lines represent the Gaussian fits to our [Fe/H] values and those from \citet{genovali2015}, respectively. `M' is the mean value of metallicity for the sample of $N$ stars.} 
\label{fig:hist_feh}
\end{figure}

Our homogeneous [Fe/H] values for 41 stars are in excellent agreement with those provided by \citet{luck2018} for common Cepheids. Therefore, we adopted their [Fe/H] values for the remaining 33 stars in our sample. The small mean offset ($-0.02$ dex) was applied to \citet{luck2018} stars that complement our sample to construct a homogenized sample of [Fe/H] values for all 75 MW Cepheid standards. Fig.~\ref{fig:hist_feh} displays the histograms of corrected [Fe/H] values from \citet{riess2021} and our homogenized metallicities ([Fe/H]$_\textrm{ESP}$). It is evident that our new [Fe/H] values for these Cepheids are on average $0.05$~dex smaller than previously adopted values from \citet{genovali2015} sample. However, our average [Fe/H] values are also in excellent agreement with the mean metallicity of Cepheids in the local region \citep[$\langle\textrm{[Fe/H]}\rangle\sim0.05$~dex,][]{luck2018}. Recently, \citet{groenewegen2023} also showed that the flux-weighted gravities from spectral energy distributions are in good agreement with those determined from high-resolution spectra of Cepheids in \citet{luck2018} scale, but do not agree with most other literature measurements.

\begin{deluxetable*}{rrrrrrrrrrrrrrr}
\tablecaption{Multiband photometric mean magnitudes of MW Cepheid variables. \label{tbl:phot}}
\tabletypesize{\footnotesize}
\tablewidth{0pt}
\tablehead{\colhead{ID}  & \colhead{\bf $\log P$} & \colhead{\bf $\overline{\omega}$}  & \colhead{\bf $G$} & \colhead{\bf $G_{BP}$}   &  \colhead{\bf $G_{RP}$}  & \colhead{\bf $V$} & \colhead{\bf $I$}  & \colhead{\bf $J$}  & \colhead{\bf $H$} & \colhead{\bf $K_s$} & \colhead{\bf $W^H_m$}& \colhead{\bf $E_{BV}$} & \colhead{[Fe/H]} & Ref.$^f$ \\ 
    &   days    &   mas & \multicolumn{10}{c}{mag} & dex &
}
\startdata
        AA Gem&    1.053&    0.311&    9.363 &    9.956&    8.620&    9.714 &   8.547$^c$ &   7.647 &    7.191&    7.086&    6.860&    0.345&$   -0.150$& ESP\\
             	&         &    0.019&    0.074&    0.082&    0.075&    0.012&    0.030&    0.010&    0.010&    0.010&    0.023&    0.036&    0.070& \\
        AD Pup&    1.133&    0.254&    9.541 &   10.125&    8.786&    9.826 &   8.672 &   7.723$^d$ &    7.341&    7.144&    7.011&    0.363&$   -0.130$& ESP\\
             	&         &    0.018&    0.053&    0.121&    0.088&    0.014&    0.012&    0.030&    0.030&    0.030&    0.024&    0.020&    0.080& \\
        AQ Car&    0.990&    0.361&    8.586 &    9.070&    7.937&    8.864 &   7.896 &   7.164$^d$ &    6.811&    6.573&    6.373&    0.168&$   -0.052$& L18\\
             	&         &    0.017&    0.044&    0.052&    0.055&    0.012&    0.011&    0.030&    0.030&    0.030&    0.011&    0.013&    0.100& \\
        AQ Pup&    1.479&    0.294&    8.127 &    8.974&    7.250&    8.625 &   7.120 &   6.139 &    5.522&    5.336&    4.859&    0.531&$   -0.170$& ESP\\
             	&         &    0.025&    0.104&    0.103&    0.102&    0.015&    0.012&    0.012&    0.011&    0.011&    0.016&    0.017&    0.080& \\
        BK Aur&    0.903&    0.426&    9.071 &    9.678&    8.314&    9.472 &   8.242$^c$ &   7.305 &    6.871&    6.748&    6.539&    0.393&$    0.080$& ESP\\
             	&         &    0.017&    0.047&    0.048&    0.037&    0.012&    0.030&    0.011&    0.010&    0.010&    0.029&    0.026&    0.090& \\
  \hline
\enddata
\tablecomments{The star ID, periods ($\log P$), and the Wesenheit magnitudes ($W^H_m$) in {\it HST} filters are from \citet{riess2021}. The parallaxes include zero-point corrections from \citet{lindegren2021}. The reddening values were adopted from \citet{groenewegen2018}. The {\it Gaia} photometric magnitudes were taken from \citet{ripepi2022} and $VIJHK_s$-band magnitudes were adopted from \citet{bhardwaj2015}, unless specified.\\
$^a$The {\it Gaia} magnitudes were taken from \citet{vallenari2023}.\\
$^{b/c}$The $V/I$-band magnitudes were derived using {\it Gaia} to Johnson-Cousin photometric system based on transformation from \citet{pancino2022}.\\
$^d$NIR ($JHK_s$) magnitudes were taken from \citet{groenewegen2018}.\\
$^e$For CR Car, NIR magnitudes, reddening, and metallicities were taken from \citet{trentin2023}.\\
$^f$The source of spectroscopic metallicities: CFHT spectra (ESP) and measurements from \citet[L18,][]{luck2018} after correcting for a mean offset of $-0.02$ dex (see text for details).\\
(This table is available in its entirety in machine-readable form.)}
\vspace{-15pt}
\end{deluxetable*}

\subsection{Multiband photometric mean-magnitudes}

We extracted {\it Gaia} photometry and astrometry for all 75 MW Cepheid standards. The intensity-weighted mean magnitudes in $G$, $G_{BP}$, and $G_{RP}$ filters for 71 Cepheids were taken from \citet{ripepi2022} based on {\it Gaia} light curves. For the remaining stars, mean magnitudes were taken from \citet{vallenari2023} source catalog. We also adopted {\it Gaia} parallaxes including the zero-point offset corrections from \citet{lindegren2021} and the distances from \citet{bailer2021}. Optical (72/65 stars in $V/I$) and near-infrared (44 stars in $JHK_s$) light curves for Cepheids were adopted from the homogenized compilation of \citet{bhardwaj2015}. Intensity-weighted mean magnitudes were derived from their Fourier-fitted light curves. When the light curves were not available, $JHK_s$ magnitudes were taken from \citet{groenewegen2018}. For the remaining $VI$-band magnitudes, we utilized photometric transformations provided by \citet{pancino2022} between {\it Gaia} and the Johnson-Kron-Cousins system. For Cepheids with available $V$ and $I$ band light curves, we found a mean offset of $-0.015\pm0.029$~mag and $0.034\pm0.030$~mag, respectively, between their intensity-weighted mean magnitudes and photometrically transformed magnitudes. An uncertainty of 0.03 mag was assigned for all magnitudes for which light curves were not available. The {\it HST} photometry ($W^H_m$ Wesenheit magnitude) for Cepheids was also adopted from \citet{riess2021}. Table~\ref{tbl:phot} lists photometric mean magnitudes and spectroscopic metallicities for all 75 MW Cepheids.

To apply reddening corrections, the color-excess, $E(B-V)$, values were taken from \citet{groenewegen2018}. 
We adopted the \citet{fitzpatrick1999} reddening law assuming $R_V = 3.1$, and corrected for extinction using the total-to-selective absorption ratios - $R_{\lambda}=2.783/1.777/0.812/0.508/0.349$ for $G$, $I$, $J$, $H$, and $K_s$, respectively \citep{breuval2022}. The photometric mean magnitudes in the $VIJHK_s$ bands were also used to construct-reddening free Wesenheit magnitudes: $W_G=G-1.9(BP-RP)$ from \citet{ripepi2019}, and ~$W_V=I-1.387(V-I),~W_{JK_s}=K-0.735(J-K_s),~W_{VK_s}=K_s-0.127(V-K_s),$ following \citet{breuval2022}. 

\section{Metallicity dependence of Period-Luminosity relations}
\label{sec:plrs}

\subsection{Linear regression fits}

The photometric properties, parallaxes, and spectroscopic metallicities listed in Table~\ref{tbl:phot} were used to derive PLZ and PWZ relations for Cepheids at multiple wavelengths. The extinction-corrected absolute magnitude or the Wesenheit magnitude ($M_{\lambda_i}$) of the $i^\textrm{th}$ Cepheid with period ($P_i$) and metallicity ([Fe/H]$_i$) at a given  wavelength ($\lambda$) is defined as:
\begin{eqnarray}
    M_{\lambda_{i}} = \alpha_\lambda + \beta_\lambda (\log P_{i} - 1) + \gamma_\lambda \textrm{[Fe/H]}_{i},
    \label{eq:plzr_lin}
\end{eqnarray}

\begin{figure*}
\includegraphics[width=0.95\textwidth]{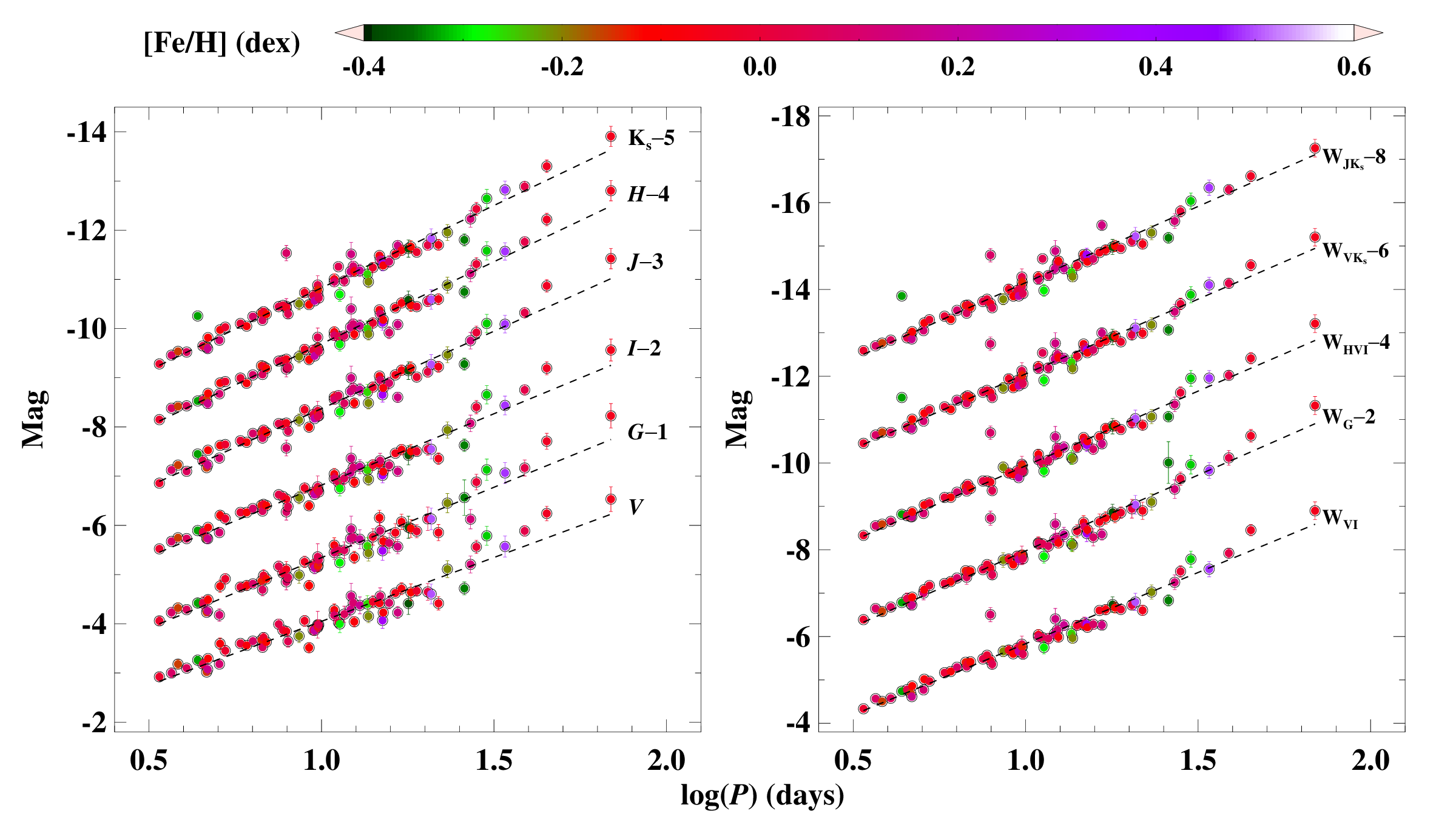}\\
\caption{Period-Luminosity-Metallicity (left) and Period-Wesenheit-Metallicity (right) relations for 68 MW Cepheids at multiple bands. The dashed lines represent the best-fitting linear regression.} 
\label{fig:linfit}
\end{figure*}

\noindent where the absolute or Wesenheit magnitudes were obtained using the distances provided by \citet{bailer2021}, which include the parallax zero-point correction by \citet{lindegren2021}. These distances were also corrected for a parallax over-correction of d$\overline{\omega}=-14~\mu$as \citep{riess2021} using the approximation $dr = -r^2 d\overline{\omega}$ adopted in \citet{breuval2022}. The $\alpha$, $\beta$, and $\gamma$ represent the zero-point, slope, and the metallicity coefficient, respectively. 

Following \citet{riess2021}, we excluded six Cepheids - DL Cas, RW Cam, SY Nor, AD Pup, U Aql, and SV Per, which exhibit values greater than 12.5 in the Gaia astrometric quality flag ({\it astrometric\_gof\_al}). These Cepheids have poor parallaxes and are all known as members of binary systems: DL Cas \citep{gieren1994}, RW Cam and SV Per \citep{evans1995, riess2018a}, SY Nor \citep{kervella2019}, AD Pup \citep{szabados2013}, U Aql \citep{evans1992, gallenne2019}.
The faintest Cepheid in our sample, CY Aur, was also excluded since its mean $G$-band magnitude falls near a sharp inflection point of larger parallax correction \citep{riess2021}. For this sample of 68 Cepheids, we fitted a linear regression in the form of equation~(\ref{eq:plzr_lin}). Given that the parallax uncertainties for these 68 Cepheids are relatively small ($<12\%$) and they are nearby ($<7$~kpc), we used distances from \citet{bailer2021} instead of parallaxes for the initial analysis to investigate PLZ/PWZ relations. We applied an iterative outlier clipping method which removed a Cepheid with the largest residual in each iteration until all stars were within $\pm2.5\sigma$ scatter. We adopted the bootstrapping procedure to create $10^4$ random realizations of the fits and determine errors.

Fig.~\ref{fig:linfit} displays PLZ and PWZ relations for Cepheids at multiple wavelengths. The figure shows that the seven longest-period Cepheids ($\log P > 1.42$) in our sample are brighter than the best-fitting linear regression. This was also seen for the $H$-band PL relation in \citet[][Figure 9]{reyes2023} and $W^H_m$ PW relation in \citet[][Figure 3]{riess2022a} for these Cepheid standards. The mean metallicities of these long-period Cepheids is $+0.07$ dex with a scatter of $0.17$~dex.
These Cepheids have $\textrm{[Fe/H]}>0$ except for 
AQ Pup ($\textrm{[Fe/H]}=-0.17\pm0.08$~dex). The lack of metal-poor stars at the long-period end in our sample can impact the slopes of the derived PLZ/PWZ relations.
Although the non-linearity of Cepheid PL relations has been extensively investigated previously \citep[e.g.,][]{bhardwaj2016b}, we only fit a single-slope model due to lower statistics at the long-period end. It is also evident that one of the Cepheids RX Cam ($\log P=0.89$) is an outlier in all PWZ relations. Hence this star is excluded from the subsequent analysis.

\begin{deluxetable*}{rcccccccccc}
\huge
\tablecaption{Multiband MW Cepheid period-luminosity-metallicity and period-Wesenheit-metallicity relations. \label{tbl:plzr}}
\tabletypesize{\footnotesize}
\tablewidth{0pt}
\tablehead{\colhead{Band}  & \colhead{$\alpha$}  &  \colhead{$\beta$}& \colhead{$\gamma$} & \colhead{$\sigma$}   & \colhead{$N_\textrm{f}$} & \colhead{$\chi^2_\textrm{dof}$} &  $\alpha_\textrm{LMC}$   & $\beta_\textrm{LMC}$  &   $\alpha_\textrm{MW}$    & $\gamma_f$
}
\startdata
& \multicolumn{6}{c}{PLZ/PWZ relations in the form of eq.~(\ref{eq:plzr_lin})} & \multicolumn{4}{c}{ABL fits with fixed slope in the form of eq.~(\ref{eq:plzr_abl})}\\
\cline{2-6} \cline{8-11}
              $V$ &$   -4.01\pm0.04   $&$   -2.26\pm0.08   $&$   -0.21\pm0.20   $&    0.17&  63&    1.05&$   -3.19\pm0.01   $&$   -2.71\pm0.03   $&$   -3.29\pm0.04   $&$   -0.22\pm0.07   $\\
              $G$ &$   -4.31\pm0.05   $&$   -2.67\pm0.09   $&$   -0.14\pm0.24   $&    0.19&  65&    0.95&$   -3.33\pm0.01   $&$   -2.82\pm0.03   $&$   -3.52\pm0.05   $&$   -0.40\pm0.10   $\\
              $I$ &$   -4.78\pm0.04   $&$   -2.57\pm0.07   $&$   -0.28\pm0.17   $&    0.13&  63&    0.95&$   -3.85\pm0.01   $&$   -2.95\pm0.02   $&$   -3.96\pm0.04   $&$   -0.23\pm0.07   $\\
              $J$ &$   -5.36\pm0.04   $&$   -3.00\pm0.06   $&$   -0.30\pm0.17   $&    0.14&  65&    1.02&$   -4.33\pm0.01   $&$   -3.10\pm0.01   $&$   -4.46\pm0.04   $&$   -0.28\pm0.07   $\\
              $H$ &$   -5.69\pm0.03   $&$   -3.27\pm0.06   $&$   -0.37\pm0.18   $&    0.16&  68&    0.97&$   -4.68\pm0.01   $&$   -3.16\pm0.01   $&$   -4.80\pm0.03   $&$   -0.27\pm0.08   $\\
            $K_s$ &$   -5.80\pm0.04   $&$   -3.32\pm0.06   $&$   -0.43\pm0.18   $&    0.15&  67&    1.02&$   -4.71\pm0.01   $&$   -3.22\pm0.01   $&$   -4.86\pm0.04   $&$   -0.33\pm0.07   $\\
    \hline
         $W_{VI}$ &$   -5.81\pm0.04   $&$   -3.08\pm0.06   $&$   -0.36\pm0.16   $&    0.13&  65&    1.03&$   -4.77\pm0.01   $&$   -3.29\pm0.01   $&$   -4.87\pm0.03   $&$   -0.21\pm0.07   $\\
          $W_{G}$ &$   -5.94\pm0.03   $&$   -3.21\pm0.07   $&$   -0.33\pm0.16   $&    0.13&  64&    1.04&$   -4.79\pm0.01   $&$   -3.34\pm0.01   $&$   -4.98\pm0.03   $&$   -0.43\pm0.06   $\\
        $W^H_{m}$ &$   -5.92\pm0.03   $&$   -3.27\pm0.05   $&$   -0.24\pm0.15   $&    0.12&  64&    1.04&$   -4.82\pm0.02   $&$   -3.31\pm0.04   $&$   -4.95\pm0.03   $&$   -0.31\pm0.07   $\\
       $W_{VK_s}$ &$   -6.02\pm0.04   $&$   -3.40\pm0.06   $&$   -0.46\pm0.18   $&    0.14&  65&    0.99&$   -4.91\pm0.01   $&$   -3.25\pm0.01   $&$   -5.07\pm0.04   $&$   -0.37\pm0.08   $\\
       $W_{JK_s}$ &$   -6.12\pm0.04   $&$   -3.46\pm0.06   $&$   -0.45\pm0.19   $&    0.15&  66&    0.99&$   -4.99\pm0.01   $&$   -3.32\pm0.01   $&$   -5.16\pm0.04   $&$   -0.37\pm0.08   $\\
  \hline
\enddata
\tablecomments{The zero-point ($\alpha$) at $\log P=1.0$ days, slope ($\beta$), metallicity coefficient ($\gamma$), dispersion ($\sigma$), and the number of stars ($N_f$) of the final PLZ/PWZ fits of eq.~(\ref{eq:plzr_lin}) are tabulated in the left columns. The coefficients of the ABL fits ($\alpha_\textrm{LMC},~\beta_{MW}$) with a fixed slope for Cepheids in the LMC are shown in the columns on the right. The metallicity coefficient ($\gamma_f$) was determined by comparing the intercepts (at $\log P=0.7$ days) of the PL relations for Cepheids in the MW and LMC as a function of their mean-metallicities.}
\vspace{-15pt}
\end{deluxetable*}

Table~\ref{tbl:plzr} lists the coefficients of the best-fitting linear regression for each PLZ/PWZ relation. The slope and the zero-points of these relations are well-constrained, but the metallicity coefficients exhibit large uncertainties. Note that the metallicity values in our sample span $-0.21$ to $0.39$~dex, covering a total range of $\Delta$[Fe/H]$=0.6$~dex. However, the majority of Cepheids have [Fe/H] values between $-0.1$ and $0.2$~dex as seen in Fig.~\ref{fig:hist_feh}. Given that the median uncertainty on these spectroscopic measurements is of the order of $0.1$~dex, this smaller range of metallicities limits the accurate and precise determination of the metallicity coefficient of the PLZ/PWZ relations. Nevertheless, the metallicity coefficients at all wavelengths exhibit a negative sign implying that the metal-rich Cepheids are brighter. The metallicity coefficients vary between $-0.14$ to $-0.46$~mag/dex, a range that is in agreement with recent empirical studies in the literature \citep[e.g.,][]{ripepi2021, breuval2022}. Our $\gamma$ values for $K_s$, $W_{JK_s}$, and $W_{JK_s}$ PLZ/PWZ relations for MW Cepheids are in better agreement with the value of $\sim-0.45$~mag/dex obtained by \citet{ripepi2021} than that of $\sim-0.32$~mag/dex in \citet{breuval2022}. This is not surprising since our approach of using individual metallicities of MW Cepheids is similar to the work of \citet{ripepi2021}. The negative sign of the metallicity term of PWZ relations has also been predicted by stellar evolution \citep{anderson2016} and pulsation models \citep{desomma2022}.

\subsection{Astrometry-Based Luminosity fits and additional metallicity constraints with Cepheids in the LMC}

The varying precision of Cepheid parallaxes can bias absolute magnitude determination since the direction-dependent priors constructed from a three-dimensional model of the Galaxy in \citet{bailer2021} can dominate the distance estimates for faint and distant stars. Therefore, it is more appropriate to derive the coefficients of the PL relations in parallax space, which can be done using astrometry-based luminosities \citep[ABL,][]{arenou1999}. Our MW Cepheid sample is small and does not provide strong constraints on the metallicity coefficient as seen in Table~\ref{tbl:plzr}. We decided to include additional constraints from the Cepheids in the LMC, and define ABL as:
\begin{eqnarray}
 {\textrm{ABL}} ~&=&~ \overline{\omega}10^{0.2m_\lambda-2} \nonumber \\
 ~&=&~ 10^{0.2(\alpha_{\lambda,\textrm{MW}} + \beta_{\lambda,\textrm{LMC}} (\log P_{i} - 0.7))},
    \label{eq:plzr_abl}
\end{eqnarray}

\noindent where $m_\lambda$ is the extinction-corrected magnitude or the Wesenheit magnitude at a given wavelength for a Cepheid with period ($P_i$). The fixed LMC slope ($\beta_\textrm{LMC}$) was adopted from \citet{breuval2022} to constrain the zero-point of the MW Cepheid PL relations ($\alpha_\textrm{MW}$). The absolute zero-point was also shifted to $\log P=0.7$ to compare with $\alpha_\textrm{LMC}$ at the same period from \citet{breuval2022}. The mean metallicity of our homogenized sample is $\langle\textrm{[Fe/H]}_\textrm{ESP}\rangle = +0.04$ dex with a dispersion of 0.09~dex. A mean metallicity of $-0.41\pm0.02$ was adopted for the LMC Cepheids \citep{romaniello2022}. The metallicity coefficient was then calculated using $\alpha = \gamma_f(\textrm{[Fe/H]}-\langle\textrm{[Fe/H]}_\textrm{ESP}\rangle) + \delta$. Since we have only two intercepts (MW and LMC), measuring the zero-point at the mean-metallicity of MW Cepheids ensures that the $\delta$ is the same as $\alpha_{MW}$ in Table~\ref{tbl:plzr}. 

The last column of Table~\ref{tbl:plzr} lists the metallicity coefficients of PLZ/PWZ relations obtained using this approach. The range of $\gamma_f$ is similar to the one obtained using linear regression. However, these coefficients are determined up to three times more precisely thanks to the additional constraint on the slope using the Cepheids in the LMC. The metallicity coefficients of the PLZ/PWZ relations vary between $-0.21$~dex and $-0.43$~mag/dex for different wavelengths and are in good agreement with those from \citet{breuval2022}. The adopted reddening values for our sample of Cepheids are different than those adopted in \citet{breuval2022}, which can also impact the determined zero-points and the metallicity coefficients of PL relations. The $\gamma_f$ values for the most PWZ relations are now smaller than those from \citet{ripepi2021} but within $1\sigma$ errors.
For this small sample of Cepheids, it can be noted that the metallicity coefficient varies depending on the adopted methodology of using individual abundances of MW Cepheids only or using Galactic and Magellanic Cloud Cepheids together assuming their mean-metallicities for the quantification of metallicity coefficient, but appear statistically consistent.

\subsection{Implications for the Hubble constant}
 
Our new spectroscopic metallicities for MW Cepheid standards were also used to derive the $W^H_m$ PWZ relation using their {\it HST} photometry from \citet{riess2021}. From Table~\ref{tbl:plzr}, the zero-point, slope, and the metallicity coefficient of $W^H_m$ PWZ relation are: $-5.92\pm0.03$~mag, $-3.27\pm0.05$~mag/dex, and $-0.24\pm0.15$ mag/dex, which agree within $0.5\sigma$ with the results of free-parameter fits in \citet[][their 4-parameter solution in Table 2]{riess2021}. 
The metallicity coefficients differ by $0.04$ mag/dex, but the uncertainty on this parameter is largely due to low statistics and limited metallicity range. This $\gamma$ parameter is smaller (in absolute sense) than $-0.37\pm0.09$~mag/dex determined by \citet{ripepi2021} using a larger sample of MW Cepheids. Comparing the $\gamma$ value obtained using MW and LMC intercepts for $W^H_m$ PWZ relation, our value of $-0.31\pm0.07$~mag/dex is also in agreement with a value of $-0.28\pm0.08$~mag/dex determined by \citet{breuval2022} using a similar approach. 

To investigate possible impact of our new spectroscopic metallicities on the local $H_0$ determination, we replaced our [Fe/H] values with those adopted in the SH0ES project \citep{riess2022}. The additional constraints on the coefficients of the PL relation come from the two independent anchors - the LMC and NGC 4258. We consider only two anchor variants of $H_0$ for our purpose: (1) MW + LMC + NGC 4258; and (2) MW. These two calibrations will allow us to explore possible variations in the metallicity coefficient and corresponding systematics due to differences in the mean metallicities of the calibrators and the target SNe host galaxies. 

Adopting all three anchors and using our [Fe/H] values in the global-matrix formalism of the SH0ES project \citep[see Section 2.1 of ][]{riess2022,riess2022a}, we obtained a metallicity coefficient of $\gamma=-0.23\pm 0.04$~mag/dex and an $H_0 = 72.9\pm 1.0$ km~s$^{-1}$~Mpc$^{-1}$. The difference in $H_0$ is 0.1 km/s/Mpc as compared to \citet{riess2022a}. In the case of only the MW as an anchor, these parameters are $\gamma=-0.21\pm 0.05$~mag/dex and an $H_0 = 72.7\pm 1.3$ km~s$^{-1}$~Mpc$^{-1}$, a difference in $H_0$ of 0.2 km/s/Mpc as compared to \citet{riess2022a}. This suggests that the systematic variations in the metallicities of Galactic Cepheid standards due to adopted metallicity scale produce changes in $H_0$ that are much smaller than the statistical uncertainty in the local $H_0$ determinations \citep{riess2022}.

\section{Summary}
\label{sec:discuss}

The Milky Way Cepheids photometrically observed with {\it HST} play a crucial role in calibrating the anchor PL relation in the same photometric system as the target Cepheids in Supernovae host galaxies. We presented new high-resolution, high signal-to-noise spectra for 42 of these 75 MW Cepheid standards using ESPaDOnS instrument at CFHT. A homogenized sample of spectroscopic [Fe/H] values for all 75 Cepheids was obtained for the calibration of Galactic Cepheid PLZ and PWZ relations for the distance scale applications. Our spectroscopic metallicities are in good agreement with most literature measurements in different metallicity scales, but are systematically smaller by $0.05$~dex than previously adopted metallicities for these 75 Cepheids in the local determination of $H_0$. The mean metallicity of our sample ${\langle\textrm{[Fe/H]}\rangle} = 0.04$~dex agrees well with the mean metallicity of Cepheids in the local region \citep{luck2018}.

The {\it Gaia} parallaxes together with the literature photometry and new spectroscopic metallicities were used to derive PLZ and PWZ relations for MW Cepheids at multiple wavelengths. The low statistics and a relatively limited range of metallicity ($\Delta \textrm{[Fe/H]}=0.6$~dex) impacted the precision of the metallicity coefficients of the PLZ/PWZ relations. Adding Cepheids in the LMC to provide additional constraints, we improve the accuracy and the precision of metallicity coefficients of the PLZ and PWZ relations. These metallicity coefficients are in good agreement with recent determinations in the literature. Moreover, we explored the differences in the metallicity coefficient of a given PLZ/PWZ relation obtained by fitting individual metallicities of MW Cepheids or comparing the intercepts of the PL/PW relations of MW and LMC Cepheids as a function of their mean metallicities. For most cases, comparing the intercepts of PL/PW relations for Cepheids in the MW and LMC gives a smaller metallicity coefficient presumably due to a narrow metallicity range ($\Delta \textrm{[Fe/H]}=0.46$~dex). Since precise geometric distances to nearby galaxies are limited to the Magellanic Clouds, it is imperative to increase the metallicity range of MW Cepheids to better constrain their PLZ/PWZ relations \citep[e.g. in][]{ripepi2021, trentin2023}. 

However, the metallicity coefficient of $W^H_m$ Wesenheit relation in {\it HST} filters does not show any statistically significant variation with our new spectroscopic metallicities. \citet{riess2022} showed that the $H_0$ values do not exhibit any correlation with metallicities because the average abundances of Cepheids in the anchor and target SNe host galaxies span the same range. Our new average [Fe/H] value ($+0.04$~dex) for MW Cepheids is smaller than the previously adopted mean metallicity ($+0.09$~dex). For a metallicity term of $\sim-0.2$ mag/dex, this would translate into a 0.01 mag difference in distance moduli measurements when the MW is the sole anchor, or less than half of this for the combination of all three anchors.  Incorporating these new metallicity measurements in the latest SH0ES distance ladder formalism \citep{riess2022}, a small difference in $H_0$ of $0.2$ km/s/Mpc was found for the MW as the only anchor case. If all three anchors are used as calibrators, which is the baseline solution for the $H_0$ determination, the new metallicities result in a little variation of $0.1$ km/s/Mpc in the $H_0$ value. 
Thus MW Cepheid metallicities in different metallicity scales make a small difference in $H_0$ relative to the size of the Hubble Tension.

\vspace{-10pt}
\acknowledgements
We thank the anonymous referee for useful comments that helped improve the manuscript.
This project has received funding from the European Union’s Horizon 2020 research and innovation programme under the Marie Skłodowska-Curie grant agreement No. 886298. This research was supported by the Munich Institute for Astro-, Particle and BioPhysics (MIAPbP) which is funded by the Deutsche Forschungsgemeinschaft (DFG, German Research Foundation) under Germany´s Excellence Strategy – EXC-2094 – 390783311.
CCN thanks the funding from the National Science and Technology Council (Taiwan) under the contract 109-2112-M-008-014-MY3.
Access to the CFHT was made possible by the Institute of Astronomy and Astrophysics, Academia Sinica.

\facility{CFHT-ESPaDOnS}
\software{\texttt{The IDL Astronomy User's Library} \citep{landsman1993},
\texttt{Astropy} \citep{astropy2013, astropy2018, astropy2022}, Libre-ESpRIT \citep{danoti1997}}

\bibliographystyle{aasjournal}
\bibliography{mybib_final.bib}

\begin{thebibliography}{}
\expandafter\ifx\csname natexlab\endcsname\relax\def\natexlab#1{#1}\fi
\providecommand{\url}[1]{\href{#1}{#1}}
\providecommand{\dodoi}[1]{doi:~\href{http://doi.org/#1}{\nolinkurl{#1}}}
\providecommand{\doeprint}[1]{\href{http://ascl.net/#1}{\nolinkurl{http://ascl.net/#1}}}
\providecommand{\doarXiv}[1]{\href{https://arxiv.org/abs/#1}{\nolinkurl{https://arxiv.org/abs/#1}}}

\bibitem[{{Acharova} {et~al.}(2012){Acharova}, {Mishurov}, \&
  {Kovtyukh}}]{acharova2012}
{Acharova}, I.~A., {Mishurov}, Y.~N., \& {Kovtyukh}, V.~V. 2012, MNRAS, 420,
  1590, \dodoi{10.1111/j.1365-2966.2011.20161.x}

\bibitem[{{Anderson} {et~al.}(2016){Anderson}, {Saio}, {Ekstr{\"o}m}, {Georgy},
  \& {Meynet}}]{anderson2016}
{Anderson}, R.~I., {Saio}, H., {Ekstr{\"o}m}, S., {Georgy}, C., \& {Meynet}, G.
  2016, A\&A, 591, A8, \dodoi{10.1051/0004-6361/201528031}

\bibitem[{{Arenou} \& {Luri}(1999)}]{arenou1999}
{Arenou}, F., \& {Luri}, X. 1999, in Astronomical Society of the Pacific
  Conference Series, Vol. 167, Harmonizing Cosmic Distance Scales in a
  Post-HIPPARCOS Era, ed. D.~{Egret} \& A.~{Heck}, 13--32,
  \dodoi{10.48550/arXiv.astro-ph/9812094}

\bibitem[{{Astropy Collaboration} {et~al.}(2013){Astropy Collaboration},
  {Robitaille}, {Tollerud}, {Greenfield}, {Droettboom}, {Bray}, {Aldcroft},
  {Davis}, {Ginsburg}, {Price-Whelan}, {Kerzendorf}, {Conley}, {Crighton},
  {Barbary}, {Muna}, {Ferguson}, {Grollier}, {Parikh}, {Nair}, {Unther},
  {Deil}, {Woillez}, {Conseil}, {Kramer}, {Turner}, {Singer}, {Fox}, {Weaver},
  {Zabalza}, {Edwards}, {Azalee Bostroem}, {Burke}, {Casey}, {Crawford},
  {Dencheva}, {Ely}, {Jenness}, {Labrie}, {Lim}, {Pierfederici}, {Pontzen},
  {Ptak}, {Refsdal}, {Servillat}, \& {Streicher}}]{astropy2013}
{Astropy Collaboration}, {Robitaille}, T.~P., {Tollerud}, E.~J., {et~al.} 2013,
  \aap, 558, A33, \dodoi{10.1051/0004-6361/201322068}

\bibitem[{{Astropy Collaboration} {et~al.}(2018){Astropy Collaboration},
  {Price-Whelan}, {Sip{\H{o}}cz}, {G{\"u}nther}, {Lim}, {Crawford}, {Conseil},
  {Shupe}, {Craig}, {Dencheva}, {Ginsburg}, {VanderPlas}, {Bradley},
  {P{\'e}rez-Su{\'a}rez}, {de Val-Borro}, {Aldcroft}, {Cruz}, {Robitaille},
  {Tollerud}, {Ardelean}, {Babej}, {Bach}, {Bachetti}, {Bakanov}, {Bamford},
  {Barentsen}, {Barmby}, {Baumbach}, {Berry}, {Biscani}, {Boquien}, {Bostroem},
  {Bouma}, {Brammer}, {Bray}, {Breytenbach}, {Buddelmeijer}, {Burke},
  {Calderone}, {Cano Rodr{\'\i}guez}, {Cara}, {Cardoso}, {Cheedella}, {Copin},
  {Corrales}, {Crichton}, {D'Avella}, {Deil}, {Depagne}, {Dietrich}, {Donath},
  {Droettboom}, {Earl}, {Erben}, {Fabbro}, {Ferreira}, {Finethy}, {Fox},
  {Garrison}, {Gibbons}, {Goldstein}, {Gommers}, {Greco}, {Greenfield},
  {Groener}, {Grollier}, {Hagen}, {Hirst}, {Homeier}, {Horton}, {Hosseinzadeh},
  {Hu}, {Hunkeler}, {Ivezi{\'c}}, {Jain}, {Jenness}, {Kanarek}, {Kendrew},
  {Kern}, {Kerzendorf}, {Khvalko}, {King}, {Kirkby}, {Kulkarni}, {Kumar},
  {Lee}, {Lenz}, {Littlefair}, {Ma}, {Macleod}, {Mastropietro}, {McCully},
  {Montagnac}, {Morris}, {Mueller}, {Mumford}, {Muna}, {Murphy}, {Nelson},
  {Nguyen}, {Ninan}, {N{\"o}the}, {Ogaz}, {Oh}, {Parejko}, {Parley}, {Pascual},
  {Patil}, {Patil}, {Plunkett}, {Prochaska}, {Rastogi}, {Reddy Janga},
  {Sabater}, {Sakurikar}, {Seifert}, {Sherbert}, {Sherwood-Taylor}, {Shih},
  {Sick}, {Silbiger}, {Singanamalla}, {Singer}, {Sladen}, {Sooley},
  {Sornarajah}, {Streicher}, {Teuben}, {Thomas}, {Tremblay}, {Turner},
  {Terr{\'o}n}, {van Kerkwijk}, {de la Vega}, {Watkins}, {Weaver}, {Whitmore},
  {Woillez}, {Zabalza}, \& {Astropy Contributors}}]{astropy2018}
{Astropy Collaboration}, {Price-Whelan}, A.~M., {Sip{\H{o}}cz}, B.~M., {et~al.}
  2018, AJ, 156, 123, \dodoi{10.3847/1538-3881/aabc4f}

\bibitem[{{Astropy Collaboration} {et~al.}(2022){Astropy Collaboration},
  {Price-Whelan}, {Lim}, {Earl}, {Starkman}, {Bradley}, {Shupe}, {Patil},
  {Corrales}, {Brasseur}, {N{\"o}the}, {Donath}, {Tollerud}, {Morris},
  {Ginsburg}, {Vaher}, {Weaver}, {Tocknell}, {Jamieson}, {van Kerkwijk},
  {Robitaille}, {Merry}, {Bachetti}, {G{\"u}nther}, {Aldcroft},
  {Alvarado-Montes}, {Archibald}, {B{\'o}di}, {Bapat}, {Barentsen},
  {Baz{\'a}n}, {Biswas}, {Boquien}, {Burke}, {Cara}, {Cara}, {Conroy},
  {Conseil}, {Craig}, {Cross}, {Cruz}, {D'Eugenio}, {Dencheva}, {Devillepoix},
  {Dietrich}, {Eigenbrot}, {Erben}, {Ferreira}, {Foreman-Mackey}, {Fox},
  {Freij}, {Garg}, {Geda}, {Glattly}, {Gondhalekar}, {Gordon}, {Grant},
  {Greenfield}, {Groener}, {Guest}, {Gurovich}, {Handberg}, {Hart},
  {Hatfield-Dodds}, {Homeier}, {Hosseinzadeh}, {Jenness}, {Jones}, {Joseph},
  {Kalmbach}, {Karamehmetoglu}, {Ka{\l}uszy{\'n}ski}, {Kelley}, {Kern},
  {Kerzendorf}, {Koch}, {Kulumani}, {Lee}, {Ly}, {Ma}, {MacBride}, {Maljaars},
  {Muna}, {Murphy}, {Norman}, {O'Steen}, {Oman}, {Pacifici}, {Pascual},
  {Pascual-Granado}, {Patil}, {Perren}, {Pickering}, {Rastogi}, {Roulston},
  {Ryan}, {Rykoff}, {Sabater}, {Sakurikar}, {Salgado}, {Sanghi}, {Saunders},
  {Savchenko}, {Schwardt}, {Seifert-Eckert}, {Shih}, {Jain}, {Shukla}, {Sick},
  {Simpson}, {Singanamalla}, {Singer}, {Singhal}, {Sinha}, {Sip{\H{o}}cz},
  {Spitler}, {Stansby}, {Streicher}, {{\v{S}}umak}, {Swinbank}, {Taranu},
  {Tewary}, {Tremblay}, {de Val-Borro}, {Van Kooten}, {Vasovi{\'c}}, {Verma},
  {de Miranda Cardoso}, {Williams}, {Wilson}, {Winkel}, {Wood-Vasey}, {Xue},
  {Yoachim}, {Zhang}, {Zonca}, \& {Astropy Project Contributors}}]{astropy2022}
{Astropy Collaboration}, {Price-Whelan}, A.~M., {Lim}, P.~L., {et~al.} 2022,
  ApJ, 935, 167, \dodoi{10.3847/1538-4357/ac7c74}

\bibitem[{{Bailer-Jones} {et~al.}(2021){Bailer-Jones}, {Rybizki}, {Fouesneau},
  {Demleitner}, \& {Andrae}}]{bailer2021}
{Bailer-Jones}, C.~A.~L., {Rybizki}, J., {Fouesneau}, M., {Demleitner}, M., \&
  {Andrae}, R. 2021, AJ, 161, 147, \dodoi{10.3847/1538-3881/abd806}

\bibitem[{{Bhardwaj} {et~al.}(2016){Bhardwaj}, {Kanbur}, {Macri}, {Singh},
  {Ngeow}, \& {Ishida}}]{bhardwaj2016b}
{Bhardwaj}, A., {Kanbur}, S.~M., {Macri}, L.~M., {et~al.} 2016, MNRAS, 457,
  1644, \dodoi{10.1093/mnras/stw040}

\bibitem[{{Bhardwaj} {et~al.}(2015){Bhardwaj}, {Kanbur}, {Singh}, {Macri}, \&
  {Ngeow}}]{bhardwaj2015}
{Bhardwaj}, A., {Kanbur}, S.~M., {Singh}, H.~P., {Macri}, L.~M., \& {Ngeow},
  C.-C. 2015, MNRAS, 447, 3342, \dodoi{10.1093/mnras/stu2678}

\bibitem[{{Breuval} {et~al.}(2022){Breuval}, {Riess}, {Kervella}, {Anderson},
  \& {Romaniello}}]{breuval2022}
{Breuval}, L., {Riess}, A.~G., {Kervella}, P., {Anderson}, R.~I., \&
  {Romaniello}, M. 2022, ApJ, 939, 89, \dodoi{10.3847/1538-4357/ac97e2}

\bibitem[{{Cruz Reyes} \& {Anderson}(2023)}]{reyes2023}
{Cruz Reyes}, M., \& {Anderson}, R.~I. 2023, A\&A, 672, A85,
  \dodoi{10.1051/0004-6361/202244775}

\bibitem[{{da Silva} {et~al.}(2022){da Silva}, {Crestani}, {Bono}, {Braga},
  {D'Orazi}, {Lemasle}, {Bergemann}, {Dall'Ora}, {Fiorentino},
  {Fran{\c{c}}ois}, {Groenewegen}, {Inno}, {Kovtyukh}, {Kudritzki},
  {Matsunaga}, {Monelli}, {Pietrinferni}, {Porcelli}, {Storm}, {Tantalo}, \&
  {Th{\'e}v{\'e}nin}}]{dasilva2022}
{da Silva}, R., {Crestani}, J., {Bono}, G., {et~al.} 2022, A\&A, 661, A104,
  \dodoi{10.1051/0004-6361/202142957}

\bibitem[{{De Somma} {et~al.}(2022){De Somma}, {Marconi}, {Molinaro}, {Ripepi},
  {Leccia}, \& {Musella}}]{desomma2022}
{De Somma}, G., {Marconi}, M., {Molinaro}, R., {et~al.} 2022, ApJS, 262, 25,
  \dodoi{10.3847/1538-4365/ac7f3b}

\bibitem[{{Di Valentino} {et~al.}(2021){Di Valentino}, {Mena}, {Pan},
  {Visinelli}, {Yang}, {Melchiorri}, {Mota}, {Riess}, \&
  {Silk}}]{valentino2021}
{Di Valentino}, E., {Mena}, O., {Pan}, S., {et~al.} 2021, Classical and Quantum
  Gravity, 38, 153001, \dodoi{10.1088/1361-6382/ac086d}

\bibitem[{{Donati} {et~al.}(1997){Donati}, {Semel}, {Carter}, {Rees}, \&
  {Collier Cameron}}]{danoti1997}
{Donati}, J.~F., {Semel}, M., {Carter}, B.~D., {Rees}, D.~E., \& {Collier
  Cameron}, A. 1997, MNRAS, 291, 658, \dodoi{10.1093/mnras/291.4.658}

\bibitem[{{Evans}(1992)}]{evans1992}
{Evans}, N.~R. 1992, ApJ, 389, 657, \dodoi{10.1086/171238}

\bibitem[{{Evans}(1995)}]{evans1995}
---. 1995, ApJ, 445, 393, \dodoi{10.1086/175704}

\bibitem[{{Fitzpatrick}(1999)}]{fitzpatrick1999}
{Fitzpatrick}, E.~L. 1999, PASP, 111, 63, \dodoi{10.1086/316293}

\bibitem[{{Freedman} {et~al.}(2001){Freedman}, {Madore}, {Gibson}, {Ferrarese},
  {Kelson}, {Sakai}, {Mould}, {Kennicutt}, {Ford}, {Graham}, {Huchra},
  {Hughes}, {Illingworth}, {Macri}, \& {Stetson}}]{freedman2001}
{Freedman}, W.~L., {Madore}, B.~F., {Gibson}, B.~K., {et~al.} 2001, ApJ, 553,
  47, \dodoi{10.1086/320638}

\bibitem[{{Gaia Collaboration} {et~al.}(2023){Gaia Collaboration}, {Vallenari},
  {Brown}, {Prusti}, {de Bruijne}, {Arenou}, {Babusiaux}, {Biermann},
  {Creevey}, {Ducourant}, {Evans}, {Eyer}, {Guerra}, {Hutton}, {Jordi},
  {Klioner}, {Lammers}, {Lindegren}, {Luri}, {Mignard}, {Panem}, {Pourbaix},
  {Randich}, {Sartoretti}, {Soubiran}, {Tanga}, {Walton}, {Bailer-Jones},
  {Bastian}, {Drimmel}, {Jansen}, {Katz}, {Lattanzi}, {van Leeuwen}, {Bakker},
  {Cacciari}, {Casta{\~n}eda}, {De Angeli}, {Fabricius}, {Fouesneau},
  {Fr{\'e}mat}, {Galluccio}, {Guerrier}, {Heiter}, {Masana}, {Messineo},
  {Mowlavi}, {Nicolas}, {Nienartowicz}, {Pailler}, {Panuzzo}, {Riclet}, {Roux},
  {Seabroke}, {Sordo}, {Th{\'e}venin}, {Gracia-Abril}, {Portell}, {Teyssier},
  {Altmann}, {Andrae}, {Audard}, {Bellas-Velidis}, {Benson}, {Berthier},
  {Blomme}, {Burgess}, {Busonero}, {Busso}, {C{\'a}novas}, {Carry}, {Cellino},
  {Cheek}, {Clementini}, {Damerdji}, {Davidson}, {de Teodoro}, {Nu{\~n}ez
  Campos}, {Delchambre}, {Dell'Oro}, {Esquej}, {Fern{\'a}ndez-Hern{\'a}ndez},
  {Fraile}, {Garabato}, {Garc{\'\i}a-Lario}, {Gosset}, {Haigron}, {Halbwachs},
  {Hambly}, {Harrison}, {Hern{\'a}ndez}, {Hestroffer}, {Hodgkin}, {Holl},
  {Jan{\ss}en}, {Jevardat de Fombelle}, {Jordan}, {Krone-Martins}, {Lanzafame},
  {L{\"o}ffler}, {Marchal}, {Marrese}, {Moitinho}, {Muinonen}, {Osborne},
  {Pancino}, {Pauwels}, {Recio-Blanco}, {Reyl{\'e}}, {Riello}, {Rimoldini},
  {Roegiers}, {Rybizki}, {Sarro}, {Siopis}, {Smith}, {Sozzetti}, {Utrilla},
  {van Leeuwen}, {Abbas}, {{\'A}brah{\'a}m}, {Abreu Aramburu}, {Aerts},
  {Aguado}, {Ajaj}, {Aldea-Montero}, {Altavilla}, {{\'A}lvarez}, {Alves},
  {Anders}, {Anderson}, {Anglada Varela}, {Antoja}, {Baines}, {Baker},
  {Balaguer-N{\'u}{\~n}ez}, {Balbinot}, {Balog}, {Barache}, {Barbato},
  {Barros}, {Barstow}, {Bartolom{\'e}}, {Bassilana}, {Bauchet}, {Becciani},
  {Bellazzini}, {Berihuete}, {Bernet}, {Bertone}, {Bianchi}, {Binnenfeld},
  {Blanco-Cuaresma}, {Blazere}, {Boch}, {Bombrun}, {Bossini}, {Bouquillon},
  {Bragaglia}, {Bramante}, {Breedt}, {Bressan}, {Brouillet}, {Brugaletta},
  {Bucciarelli}, {Burlacu}, {Butkevich}, {Buzzi}, {Caffau}, {Cancelliere},
  {Cantat-Gaudin}, {Carballo}, {Carlucci}, {Carnerero}, {Carrasco},
  {Casamiquela}, {Castellani}, {Castro-Ginard}, {Chaoul}, {Charlot}, {Chemin},
  {Chiaramida}, {Chiavassa}, {Chornay}, {Comoretto}, {Contursi}, {Cooper},
  {Cornez}, {Cowell}, {Crifo}, {Cropper}, {Crosta}, {Crowley}, {Dafonte},
  {Dapergolas}, {David}, {David}, {de Laverny}, {De Luise}, {De March}, {De
  Ridder}, {de Souza}, {de Torres}, {del Peloso}, {del Pozo}, {Delbo},
  {Delgado}, {Delisle}, {Demouchy}, {Dharmawardena}, {Di Matteo}, {Diakite},
  {Diener}, {Distefano}, {Dolding}, {Edvardsson}, {Enke}, {Fabre}, {Fabrizio},
  {Faigler}, {Fedorets}, {Fernique}, {Fienga}, {Figueras}, {Fournier},
  {Fouron}, {Fragkoudi}, {Gai}, {Garcia-Gutierrez}, {Garcia-Reinaldos},
  {Garc{\'\i}a-Torres}, {Garofalo}, {Gavel}, {Gavras}, {Gerlach}, {Geyer},
  {Giacobbe}, {Gilmore}, {Girona}, {Giuffrida}, {Gomel}, {Gomez},
  {Gonz{\'a}lez-N{\'u}{\~n}ez}, {Gonz{\'a}lez-Santamar{\'\i}a},
  {Gonz{\'a}lez-Vidal}, {Granvik}, {Guillout}, {Guiraud},
  {Guti{\'e}rrez-S{\'a}nchez}, {Guy}, {Hatzidimitriou}, {Hauser}, {Haywood},
  {Helmer}, {Helmi}, {Sarmiento}, {Hidalgo}, {Hilger}, {H{\l}adczuk}, {Hobbs},
  {Holland}, {Huckle}, {Jardine}, {Jasniewicz}, {Jean-Antoine Piccolo},
  {Jim{\'e}nez-Arranz}, {Jorissen}, {Juaristi Campillo}, {Julbe}, {Karbevska},
  {Kervella}, {Khanna}, {Kontizas}, {Kordopatis}, {Korn}, {K{\'o}sp{\'a}l},
  {Kostrzewa-Rutkowska}, {Kruszy{\'n}ska}, {Kun}, {Laizeau}, {Lambert},
  {Lanza}, {Lasne}, {Le Campion}, {Lebreton}, {Lebzelter}, {Leccia}, {Leclerc},
  {Lecoeur-Taibi}, {Liao}, {Licata}, {Lindstr{\o}m}, {Lister}, {Livanou},
  {Lobel}, {Lorca}, {Loup}, {Madrero Pardo}, {Magdaleno Romeo}, {Managau},
  {Mann}, {Manteiga}, {Marchant}, {Marconi}, {Marcos}, {Marcos Santos},
  {Mar{\'\i}n Pina}, {Marinoni}, {Marocco}, {Marshall}, {Martin Polo},
  {Mart{\'\i}n-Fleitas}, {Marton}, {Mary}, {Masip}, {Massari},
  {Mastrobuono-Battisti}, {Mazeh}, {McMillan}, {Messina}, {Michalik}, {Millar},
  {Mints}, {Molina}, {Molinaro}, {Moln{\'a}r}, {Monari}, {Mongui{\'o}},
  {Montegriffo}, {Montero}, {Mor}, {Mora}, {Morbidelli}, {Morel}, {Morris},
  {Muraveva}, {Murphy}, {Musella}, {Nagy}, {Noval}, {Oca{\~n}a}, {Ogden},
  {Ordenovic}, {Osinde}, {Pagani}, {Pagano}, {Palaversa}, {Palicio},
  {Pallas-Quintela}, {Panahi}, {Payne-Wardenaar}, {Pe{\~n}alosa Esteller},
  {Penttil{\"a}}, {Pichon}, {Piersimoni}, {Pineau}, {Plachy}, {Plum}, {Poggio},
  {Pr{\v{s}}a}, {Pulone}, {Racero}, {Ragaini}, {Rainer}, {Raiteri}, {Rambaux},
  {Ramos}, {Ramos-Lerate}, {Re Fiorentin}, {Regibo}, {Richards}, {Rios Diaz},
  {Ripepi}, {Riva}, {Rix}, {Rixon}, {Robichon}, {Robin}, {Robin}, {Roelens},
  {Rogues}, {Rohrbasser}, {Romero-G{\'o}mez}, {Rowell}, {Royer}, {Ruz Mieres},
  {Rybicki}, {Sadowski}, {S{\'a}ez N{\'u}{\~n}ez}, {Sagrist{\`a} Sell{\'e}s},
  {Sahlmann}, {Salguero}, {Samaras}, {Sanchez Gimenez}, {Sanna},
  {Santove{\~n}a}, {Sarasso}, {Schultheis}, {Sciacca}, {Segol}, {Segovia},
  {S{\'e}gransan}, {Semeux}, {Shahaf}, {Siddiqui}, {Siebert}, {Siltala},
  {Silvelo}, {Slezak}, {Slezak}, {Smart}, {Snaith}, {Solano}, {Solitro},
  {Souami}, {Souchay}, {Spagna}, {Spina}, {Spoto}, {Steele},
  {Steidelm{\"u}ller}, {Stephenson}, {S{\"u}veges}, {Surdej}, {Szabados},
  {Szegedi-Elek}, {Taris}, {Taylor}, {Teixeira}, {Tolomei}, {Tonello}, {Torra},
  {Torra}, {Torralba Elipe}, {Trabucchi}, {Tsounis}, {Turon}, {Ulla}, {Unger},
  {Vaillant}, {van Dillen}, {van Reeven}, {Vanel}, {Vecchiato}, {Viala},
  {Vicente}, {Voutsinas}, {Weiler}, {Wevers}, {Wyrzykowski}, {Yoldas}, {Yvard},
  {Zhao}, {Zorec}, {Zucker}, \& {Zwitter}}]{vallenari2023}
{Gaia Collaboration}, {Vallenari}, A., {Brown}, A.~G.~A., {et~al.} 2023, A\&A,
  674, A1, \dodoi{10.1051/0004-6361/202243940}

\bibitem[{{Gallenne} {et~al.}(2019){Gallenne}, {Kervella}, {Borgniet},
  {M{\'e}rand}, {Pietrzy{\'n}ski}, {Gieren}, {Monnier}, {Schaefer}, {Evans},
  {Anderson}, {Baron}, {Roettenbacher}, \& {Karczmarek}}]{gallenne2019}
{Gallenne}, A., {Kervella}, P., {Borgniet}, S., {et~al.} 2019, A\&A, 622, A164,
  \dodoi{10.1051/0004-6361/201834614}

\bibitem[{{Genovali} {et~al.}(2014){Genovali}, {Lemasle}, {Bono}, {Romaniello},
  {Fabrizio}, {Ferraro}, {Iannicola}, {Laney}, {Nonino}, {Bergemann},
  {Buonanno}, {Fran{\c{c}}ois}, {Inno}, {Kudritzki}, {Matsunaga}, {Pedicelli},
  {Primas}, \& {Th{\'e}venin}}]{genovali2014}
{Genovali}, K., {Lemasle}, B., {Bono}, G., {et~al.} 2014, A\&A, 566, A37,
  \dodoi{10.1051/0004-6361/201323198}

\bibitem[{{Genovali} {et~al.}(2015){Genovali}, {Lemasle}, {da Silva}, {Bono},
  {Fabrizio}, {Bergemann}, {Buonanno}, {Ferraro}, {Fran{\c{c}}ois},
  {Iannicola}, {Inno}, {Laney}, {Kudritzki}, {Matsunaga}, {Nonino}, {Primas},
  {Romaniello}, {Urbaneja}, \& {Th{\'e}venin}}]{genovali2015}
{Genovali}, K., {Lemasle}, B., {da Silva}, R., {et~al.} 2015, A\&A, 580, A17,
  \dodoi{10.1051/0004-6361/201525894}

\bibitem[{{Gieren} {et~al.}(1994){Gieren}, {Welch}, {Mermilliod}, {Matthews},
  \& {Hertling}}]{gieren1994}
{Gieren}, W.~P., {Welch}, D.~L., {Mermilliod}, J.-C., {Matthews}, J.~M., \&
  {Hertling}, G. 1994, AJ, 107, 2093, \dodoi{10.1086/117019}

\bibitem[{{Groenewegen} \& {Lub}(2023)}]{groenewegen2023}
{Groenewegen}, M., \& {Lub}, J. 2023, arXiv e-prints, arXiv:2307.07559,
  \dodoi{10.48550/arXiv.2307.07559}

\bibitem[{{Groenewegen}(2018)}]{groenewegen2018}
{Groenewegen}, M.~A.~T. 2018, A\&A, 619, A8,
  \dodoi{10.1051/0004-6361/201833478}

\bibitem[{{Kervella} {et~al.}(2019){Kervella}, {Gallenne}, {Evans}, {Szabados},
  {Arenou}, {M{\'e}rand}, {Nardetto}, {Gieren}, \&
  {Pietrzynski}}]{kervella2019}
{Kervella}, P., {Gallenne}, A., {Evans}, N.~R., {et~al.} 2019, A\&A, 623, A117,
  \dodoi{10.1051/0004-6361/201834211}

\bibitem[{{Kovtyukh} {et~al.}(2022){Kovtyukh}, {Lemasle}, {Bono}, {Usenko}, {da
  Silva}, {Kniazev}, {Grebel}, {Andronov}, {Shakun}, \&
  {Chinarova}}]{kovtyukh2022}
{Kovtyukh}, V., {Lemasle}, B., {Bono}, G., {et~al.} 2022, MNRAS, 510, 1894,
  \dodoi{10.1093/mnras/stab3530}

\bibitem[{{Kovtyukh} \& {Gorlova}(2000)}]{kovtyukh2000}
{Kovtyukh}, V.~V., \& {Gorlova}, N.~I. 2000, A\&A, 358, 587

\bibitem[{{Kurucz}(1993)}]{kurucz1993}
{Kurucz}, R.~L. 1993, {SYNTHE spectrum synthesis programs and line data}

\bibitem[{{Kurucz}(2005)}]{kurucz2005}
---. 2005, Memorie della Societa Astronomica Italiana Supplementi, 8, 14

\bibitem[{{Landsman}(1993)}]{landsman1993}
{Landsman}, W.~B. 1993, in ASP Conference Series, Vol.~52, Astronomical Data
  Analysis Software and Systems II, ed. R.~J. {Hanisch}, R.~J.~V. {Brissenden},
  \& J.~{Barnes}, 246

\bibitem[{{Leavitt} \& {Pickering}(1912)}]{leavitt1912}
{Leavitt}, H.~S., \& {Pickering}, E.~C. 1912, Harvard College Observatory
  Circular, 173, 1

\bibitem[{{Lindegren} {et~al.}(2021){Lindegren}, {Klioner}, {Hern{\'a}ndez},
  {Bombrun}, {Ramos-Lerate}, {Steidelm{\"u}ller}, {Bastian}, {Biermann}, {de
  Torres}, {Gerlach}, {Geyer}, {Hilger}, {Hobbs}, {Lammers}, {McMillan},
  {Stephenson}, {Casta{\~n}eda}, {Davidson}, {Fabricius}, {Gracia-Abril},
  {Portell}, {Rowell}, {Teyssier}, {Torra}, {Bartolom{\'e}}, {Clotet},
  {Garralda}, {Gonz{\'a}lez-Vidal}, {Torra}, {Abbas}, {Altmann}, {Anglada
  Varela}, {Balaguer-N{\'u}{\~n}ez}, {Balog}, {Barache}, {Becciani}, {Bernet},
  {Bertone}, {Bianchi}, {Bouquillon}, {Brown}, {Bucciarelli}, {Busonero},
  {Butkevich}, {Buzzi}, {Cancelliere}, {Carlucci}, {Charlot}, {Cioni},
  {Crosta}, {Crowley}, {del Peloso}, {del Pozo}, {Drimmel}, {Esquej}, {Fienga},
  {Fraile}, {Gai}, {Garcia-Reinaldos}, {Guerra}, {Hambly}, {Hauser},
  {Jan{\ss}en}, {Jordan}, {Kostrzewa-Rutkowska}, {Lattanzi}, {Liao}, {Licata},
  {Lister}, {L{\"o}ffler}, {Marchant}, {Masip}, {Mignard}, {Mints}, {Molina},
  {Mora}, {Morbidelli}, {Murphy}, {Pagani}, {Panuzzo}, {Pe{\~n}alosa Esteller},
  {Poggio}, {Re Fiorentin}, {Riva}, {Sagrist{\`a} Sell{\'e}s}, {Sanchez
  Gimenez}, {Sarasso}, {Sciacca}, {Siddiqui}, {Smart}, {Souami}, {Spagna},
  {Steele}, {Taris}, {Utrilla}, {van Reeven}, \& {Vecchiato}}]{lindegren2021}
{Lindegren}, L., {Klioner}, S.~A., {Hern{\'a}ndez}, J., {et~al.} 2021, A\&A,
  649, A2, \dodoi{10.1051/0004-6361/202039709}

\bibitem[{{Luck}(2018)}]{luck2018}
{Luck}, R.~E. 2018, AJ, 156, 171, \dodoi{10.3847/1538-3881/aadcac}

\bibitem[{{Luck} \& {Lambert}(2011)}]{luck2011}
{Luck}, R.~E., \& {Lambert}, D.~L. 2011, AJ, 142, 136,
  \dodoi{10.1088/0004-6256/142/4/136}

\bibitem[{{Molinaro} {et~al.}(2023){Molinaro}, {Ripepi}, {Marconi},
  {Romaniello}, {Catanzaro}, {Cusano}, {De Somma}, {Musella}, {Storm}, \&
  {Trentin}}]{molinaro2023}
{Molinaro}, R., {Ripepi}, V., {Marconi}, M., {et~al.} 2023, MNRAS, 520, 4154,
  \dodoi{10.1093/mnras/stad440}

\bibitem[{{Ngeow}(2012)}]{ngeow2012}
{Ngeow}, C.-C. 2012, ApJ, 747, 50, \dodoi{10.1088/0004-637X/747/1/50}

\bibitem[{{Owens} {et~al.}(2022){Owens}, {Freedman}, {Madore}, \&
  {Lee}}]{owens2022}
{Owens}, K.~A., {Freedman}, W.~L., {Madore}, B.~F., \& {Lee}, A.~J. 2022, ApJ,
  927, 8, \dodoi{10.3847/1538-4357/ac479e}

\bibitem[{{Pancino} {et~al.}(2022){Pancino}, {Marrese}, {Marinoni}, {Sanna},
  {Turchi}, {Tsantaki}, {Rainer}, {Altavilla}, {Monelli}, \&
  {Monaco}}]{pancino2022}
{Pancino}, E., {Marrese}, P.~M., {Marinoni}, S., {et~al.} 2022, A\&A, 664,
  A109, \dodoi{10.1051/0004-6361/202243939}

\bibitem[{{Planck Collaboration} {et~al.}(2020){Planck Collaboration},
  {Aghanim}, {Akrami}, {Ashdown}, {Aumont}, {Baccigalupi}, {Ballardini},
  {Banday}, {Barreiro}, {Bartolo}, {Basak}, {Battye}, {Benabed}, {Bernard},
  {Bersanelli}, {Bielewicz}, {Bock}, {Bond}, {Borrill}, {Bouchet}, {Boulanger},
  {Bucher}, {Burigana}, {Butler}, {Calabrese}, {Cardoso}, {Carron},
  {Challinor}, {Chiang}, {Chluba}, {Colombo}, {Combet}, {Contreras}, {Crill},
  {Cuttaia}, {de Bernardis}, {de Zotti}, {Delabrouille}, {Delouis}, {Di
  Valentino}, {Diego}, {Dor{\'e}}, {Douspis}, {Ducout}, {Dupac}, {Dusini},
  {Efstathiou}, {Elsner}, {En{\ss}lin}, {Eriksen}, {Fantaye}, {Farhang},
  {Fergusson}, {Fernandez-Cobos}, {Finelli}, {Forastieri}, {Frailis},
  {Franceschi}, {Frolov}, {Galeotta}, {Galli}, {Ganga}, {G{\'e}nova-Santos},
  {Gerbino}, {Ghosh}, {Gonz{\'a}lez-Nuevo}, {G{\'o}rski}, {Gratton},
  {Gruppuso}, {Gudmundsson}, {Hamann}, {Handley}, {Herranz}, {Hivon}, {Huang},
  {Jaffe}, {Jones}, {Karakci}, {Keih{\"a}nen}, {Keskitalo}, {Kiiveri}, {Kim},
  {Kisner}, {Knox}, {Krachmalnicoff}, {Kunz}, {Kurki-Suonio}, {Lagache},
  {Lamarre}, {Lasenby}, {Lattanzi}, {Lawrence}, {Le Jeune}, {Lemos},
  {Lesgourgues}, {Levrier}, {Lewis}, {Liguori}, {Lilje}, {Lilley}, {Lindholm},
  {L{\'o}pez-Caniego}, {Lubin}, {Ma}, {Mac{\'{\i}}as-P{\'e}rez}, {Maggio},
  {Maino}, {Mandolesi}, {Mangilli}, {Marcos-Caballero}, {Maris}, {Martin},
  {Martinelli}, {Mart{\'{\i}}nez-Gonz{\'a}lez}, {Matarrese}, {Mauri}, {McEwen},
  {Meinhold}, {Melchiorri}, {Mennella}, {Migliaccio}, {Millea}, {Mitra},
  {Miville-Desch{\^e}nes}, {Molinari}, {Montier}, {Morgante}, {Moss}, {Natoli},
  {N{\o}rgaard-Nielsen}, {Pagano}, {Paoletti}, {Partridge}, {Patanchon},
  {Peiris}, {Perrotta}, {Pettorino}, {Piacentini}, {Polastri}, {Polenta},
  {Puget}, {Rachen}, {Reinecke}, {Remazeilles}, {Renzi}, {Rocha}, {Rosset},
  {Roudier}, {Rubi{\~n}o-Mart{\'{\i}}n}, {Ruiz-Granados}, {Salvati}, {Sandri},
  {Savelainen}, {Scott}, {Shellard}, {Sirignano}, {Sirri}, {Spencer},
  {Sunyaev}, {Suur-Uski}, {Tauber}, {Tavagnacco}, {Tenti}, {Toffolatti},
  {Tomasi}, {Trombetti}, {Valenziano}, {Valiviita}, {Van Tent}, {Vibert},
  {Vielva}, {Villa}, {Vittorio}, {Wandelt}, {Wehus}, {White}, {White},
  {Zacchei}, \& {Zonca}}]{planck2020}
{Planck Collaboration}, {Aghanim}, N., {Akrami}, Y., {et~al.} 2020, A\&A, 641,
  A6, \dodoi{10.1051/0004-6361/201833910}

\bibitem[{{Riess} {et~al.}(2021){Riess}, {Casertano}, {Yuan}, {Bowers},
  {Macri}, {Zinn}, \& {Scolnic}}]{riess2021}
{Riess}, A.~G., {Casertano}, S., {Yuan}, W., {et~al.} 2021, ApJL, 908, L6,
  \dodoi{10.3847/2041-8213/abdbaf}

\bibitem[{{Riess} {et~al.}(2011){Riess}, {Macri}, {Casertano}, {Lampeitl},
  {Ferguson}, {Filippenko}, {Jha}, {Li}, \& {Chornock}}]{riess2011}
{Riess}, A.~G., {Macri}, L., {Casertano}, S., {et~al.} 2011, ApJ, 730, 119,
  \dodoi{10.1088/0004-637X/730/2/119}

\bibitem[{{Riess} {et~al.}(2016){Riess}, {Macri}, {Hoffmann}, {Scolnic},
  {Casertano}, {Filippenko}, {Tucker}, {Reid}, {Jones}, {Silverman},
  {Chornock}, {Challis}, {Yuan}, {Brown}, \& {Foley}}]{riess2016}
{Riess}, A.~G., {Macri}, L.~M., {Hoffmann}, S.~L., {et~al.} 2016, ApJ, 826, 56,
  \dodoi{10.3847/0004-637X/826/1/56}

\bibitem[{{Riess} {et~al.}(2018{\natexlab{a}}){Riess}, {Casertano}, {Yuan},
  {Macri}, {Anderson}, {MacKenty}, {Bowers}, {Clubb}, {Filippenko}, {Jones}, \&
  {Tucker}}]{riess2018}
{Riess}, A.~G., {Casertano}, S., {Yuan}, W., {et~al.} 2018{\natexlab{a}}, ApJ,
  855, 136, \dodoi{10.3847/1538-4357/aaadb7}

\bibitem[{{Riess} {et~al.}(2018{\natexlab{b}}){Riess}, {Casertano}, {Yuan},
  {Macri}, {Bucciarelli}, {Lattanzi}, {MacKenty}, {Bowers}, {Zheng},
  {Filippenko}, {Huang}, \& {Anderson}}]{riess2018a}
---. 2018{\natexlab{b}}, ApJ, 861, 126, \dodoi{10.3847/1538-4357/aac82e}

\bibitem[{{Riess} {et~al.}(2022{\natexlab{a}}){Riess}, {Yuan}, {Macri},
  {Scolnic}, {Brout}, {Casertano}, {Jones}, {Murakami}, {Anand}, {Breuval},
  {Brink}, {Filippenko}, {Hoffmann}, {Jha}, {D'arcy Kenworthy}, {Mackenty},
  {Stahl}, \& {Zheng}}]{riess2022}
{Riess}, A.~G., {Yuan}, W., {Macri}, L.~M., {et~al.} 2022{\natexlab{a}}, ApJL,
  934, L7, \dodoi{10.3847/2041-8213/ac5c5b}

\bibitem[{{Riess} {et~al.}(2022{\natexlab{b}}){Riess}, {Breuval}, {Yuan},
  {Casertano}, {Macri}, {Bowers}, {Scolnic}, {Cantat-Gaudin}, {Anderson}, \&
  {Cruz Reyes}}]{riess2022a}
{Riess}, A.~G., {Breuval}, L., {Yuan}, W., {et~al.} 2022{\natexlab{b}}, ApJ,
  938, 36, \dodoi{10.3847/1538-4357/ac8f24}

\bibitem[{{Ripepi} {et~al.}(2019){Ripepi}, {Molinaro}, {Musella}, {Marconi},
  {Leccia}, \& {Eyer}}]{ripepi2019}
{Ripepi}, V., {Molinaro}, R., {Musella}, I., {et~al.} 2019, A\&A, 625, A14,
  \dodoi{10.1051/0004-6361/201834506}

\bibitem[{{Ripepi} {et~al.}(2021){Ripepi}, {Catanzaro}, {Molinaro}, {Gatto},
  {De Somma}, {Marconi}, {Romaniello}, {Leccia}, {Musella}, {Trentin},
  {Clementini}, {Testa}, {Cusano}, \& {Storm}}]{ripepi2021}
{Ripepi}, V., {Catanzaro}, G., {Molinaro}, R., {et~al.} 2021, MNRAS, 508, 4047,
  \dodoi{10.1093/mnras/stab2460}

\bibitem[{{Ripepi} {et~al.}(2022){Ripepi}, {Clementini}, {Molinaro}, {Leccia},
  {Plachy}, {Moln{\'a}r}, {Rimoldini}, {Musella}, {Marconi}, {Garofalo},
  {Audard}, {Holl}, {Evans}, {Jevardat de Fombelle}, {Lecoeur-Taibi},
  {Marchal}, {Mowlavi}, {Muraveva}, {Nienartowicz}, {Sartoretti}, {Szabados},
  \& {Eyer}}]{ripepi2022}
{Ripepi}, V., {Clementini}, G., {Molinaro}, R., {et~al.} 2022, arXiv e-prints,
  arXiv:2206.06212, \dodoi{10.48550/arXiv.2206.06212}

\bibitem[{{Romaniello} {et~al.}(2008){Romaniello}, {Primas}, {Mottini},
  {Pedicelli}, {Lemasle}, {Bono}, {Fran{\c{c}}ois}, {Groenewegen}, \&
  {Laney}}]{romaniello2008}
{Romaniello}, M., {Primas}, F., {Mottini}, M., {et~al.} 2008, A\&A, 488, 731,
  \dodoi{10.1051/0004-6361:20065661}

\bibitem[{{Romaniello} {et~al.}(2022){Romaniello}, {Riess}, {Mancino},
  {Anderson}, {Freudling}, {Kudritzki}, {Macr{\`\i}}, {Mucciarelli}, \&
  {Yuan}}]{romaniello2022}
{Romaniello}, M., {Riess}, A., {Mancino}, S., {et~al.} 2022, A\&A, 662, C1,
  \dodoi{10.1051/0004-6361/202142441e}

\bibitem[{{Szabados} {et~al.}(2013){Szabados}, {Derekas}, {Kiss}, {Kov{\'a}cs},
  {Anderson}, {Kiss}, {Szalai}, {Sz{\'e}kely}, \&
  {Christiansen}}]{szabados2013}
{Szabados}, L., {Derekas}, A., {Kiss}, L.~L., {et~al.} 2013, MNRAS, 430, 2018,
  \dodoi{10.1093/mnras/stt027}

\bibitem[{{Trentin} {et~al.}(2023){Trentin}, {Ripepi}, {Catanzaro}, {Storm},
  {Marconi}, {De Somma}, {Testa}, \& {Musella}}]{trentin2023}
{Trentin}, E., {Ripepi}, V., {Catanzaro}, G., {et~al.} 2023, MNRAS, 519, 2331,
  \dodoi{10.1093/mnras/stac2459}

\bibitem[{{Verde} {et~al.}(2019){Verde}, {Treu}, \& {Riess}}]{verde2019}
{Verde}, L., {Treu}, T., \& {Riess}, A.~G. 2019, Nature Astronomy, 3, 891,
  \dodoi{10.1038/s41550-019-0902-0}

\end{thebibliography}

\end{document}